\documentclass[aps,pra,preprint,superscriptaddress,tightenlines,showpacs]{revtex4}

\usepackage{graphicx} 
\usepackage{amsmath}
 
\newcommand{\beq}{\begin{equation}}
\newcommand{\enq}{\end{equation}}
\begin{document}

\title{Noise correlations of the ultra-cold Fermi gas in an optical lattice}
\author{T. Paananen}
\email{Tomi.Paananen@helsinki.fi}
\affiliation{Department of Physics, University of Helsinki, 
PO Box 64, 00014 University of Helsinki,  Finland}
\author{T. K. Koponen}
\affiliation{Nanoscience Center, Department of Physics,
PO Box 35, 40014 University of Jyv\"{a}skyl\"{a}, Finland}
\affiliation{Laboratory of Physics, Helsinki University of Technology, P.O. Box 5100,
02015 HUT, Finland}
\author{P. T{\"{o}}rm{\"{a}}}
\affiliation{Nanoscience Center, Department of Physics,
PO Box 35, 40014 University of Jyv\"{a}skyl\"{a}, Finland}
\affiliation{Laboratory of Physics, Helsinki University of Technology, P.O. Box 5100,
02015 HUT, Finland}
\author{J.-P. Martikainen}
\affiliation{Department of Physics, University of Helsinki, 
PO Box 64, 00014 University of Helsinki,  Finland}
\affiliation{Nordita, Roslagstullsbacken 23, 106 91 Stockholm,
Sweden}
\date{\today}

\begin{abstract}
In this paper we study the density noise correlations of the two component Fermi gas 
in optical lattices.
Three different type of phases, the BCS-state (Bardeen, Cooper, and Schieffer), 
the FFLO-state  
(Fulde, Ferrel, Larkin, and Ovchinnikov), and BP (breach pair) state, are considered.
We show how these states differ in their noise correlations.
The noise correlations are calculated not only at zero temperature, 
but also at non-zero temperatures
paying particular attention to how much the finite temperature effects might complicate
the detection of different phases. Since one-dimensional systems have been shown to be very promising 
candidates to observe FFLO states, we apply our results also to the
computation of correlation
signals in a one-dimensional lattice. We find that the density noise correlations reveal important
information about the structure of the underlying order parameter as well as about
the quasiparticle dispersions.
\end{abstract}
\pacs{03.75.-b, 32.80.Pj, 03.65.-w}  
\maketitle

\section{Introduction}
\label{sec:intro}

Recent studies in the experiments on ultracold Fermi gases have shown great potential for 
elucidating long-standing problems in many different fields of physics related to strongly correlated 
fermions. For instance, 
in recent
experiments~\cite{Zwierlein2006a,Partridge2006a,Zwierlein2006c,Shin2006a,Partridge2006c,Shin2008a} 
spin-density imbalanced, or polarized, Fermi gases were considered. Among other things, such systems 
make it possible to study pairing with mismatched Fermi surfaces, 
potentially leading to non-BCS type pairing 
such as that appearing in FFLO-states~\cite{Fulde1964a,Larkin1964a} 
or BP-states~\cite{Sarma1963a,Liu2003a} 
(Sarma-states). These possibilities have been considered 
extensively in condensed-matter, nuclear, and high-energy physics~\cite{Casalbuoni2004a}. 

In optical lattices it is possible to study  many different physical problems with close analogs 
in the field of solid state physics. However, in contrast to many solid state systems,
ultra-cold gases in optical lattices provide a very clean environment. In other words, 
these systems have very 
few imperfections and if imperfections are of interest, they can be introduced
in a more controlled manner.
In addition, since optical lattices are made with lasers, the lattice geometry is easy to 
vary~\cite{Orzel2001a,Greiner2001b,Burger2001a,Hadzibabic2004a} by changing
the properties of the intersecting laser beams.
For these reasons optical lattices enable one to investigate 
various quantum many-body physics problems, such as those related to
Mott insulators, phase coherence, and superfluidity.
Indeed, the possibility of a superfluid alkali atom Fermi gas in an optical lattice 
has been recently studied both 
theoretically~\cite{Hofstetter2002a,Orso2005a,Pitaevskii2005a,Iskin2007a,Koponen2006a,Koponen2006b,Koponen2007a}, 
as well as experimentally~\cite{Chin2006a}. Furthermore, Feshbach resonances, molecules of fermionic atoms, and p-wave interactions
in a lower dimensional fermionic systems have been studied using optical lattices~\cite{Kohl2005a,Stoferle2005a,Guenter2005a}.

Density-density correlations tell us how strongly the atomic densities at different positions
are correlated.
Often it is useful to focus on the deviation of the atomic density from its mean value
and subtract the background away.
In such cases one studies noise correlations i.e how the density
fluctuations at different positions are correlated. 
Measuring noise correlations is 
a promising way to observe different phases in an optical lattice since
while densities can be very similar for different phases, 
the noise correlations can still be very different. As an example, one can mention the famous 
phase transition between the superfluid Bose gas and the Mott insulator~\cite{Altman2004a,Foelling2005a}. 
In the Mott insulator phase one can see the Bragg peaks 
in the noise correlations, but the noise correlations 
vanish for a Bose-Einstein condensate. Experimentally noise correlations have been used 
as an indicator for different phases in optical lattices for 
bosons~\cite{Foelling2005a}  and
to observe fermionic anti-bunching for ideal fermions~\cite{Rom2006a}.
Also, noise correlations have been used to detect pairing correlations in 
an interacting Fermi gas in a harmonic trap~\cite{Greiner2005b}.

In this paper we study  noise correlations in a cloud of ultra-cold two component Fermi gas
at finite temperatures.
We are motivated by the fact that
via noise correlations one can see subtle correlation effects which are
not visible in the lower order correlations functions. Not only are the noise correlation
signals very different for a superfluid and a normal Fermi gas, but
noise correlations also differ between different types of paired states. 
While we compute density-density correlations also at zero temperature, 
special attention is payed to finite temperature calculations, since 
finite temperature effects
have not been extensively discussed before. We find that although finite temperature effects
do smooth out some sharp features present in zero temperature calculations,
qualitative and clear quantitative differences between phases
can still exist at higher temperatures. Also, we show how regions of
gapless excitations are reflected in the density-density correlations.
For instance, gapless excitations show up
as absence of correlations for certain momenta. Moreover, they are also
reflected in the value of the maximum correlation peak height, this
quantitative signature persists at finite temperature as well.

Reducing the dimensionality of the system from three dimensions into one has been
shown to favor the FFLO-type modulated order parameters
in free space~\cite{Machida1984a,Liu2007a,Parish2007a} as well as in a lattice~\cite{Koponen2007b}.
This effect of dimensionality, together with the general
tendency of a lattice geometry to favor FFLO state due to nesting of the
Fermi surfaces~\cite{Koponen2007a}, makes one dimensional optical lattices
a promising system to study the FFLO state.
For this reason, we apply our general results on density-density correlations
also to the one-dimensional (1D) system. Very recently a similar one-dimensional problem was also
discussed at zero temperature using the DMRG algorithm~\cite{Luscher2007a}.
Here we compute the density-density correlations of the one-dimensional system
at representative points of the  finite temperature phase diagram of a polarized system.
The noise correlations in a one-dimensional system turn out to show differences between different 
states which are straightforward to interpret and contain useful information
on the underlying pair-wavefunction as well as on the structure of the quasiparticle
dispersions.

This paper is organized as follows. In Sec.~\ref{sec:sysham}
we discuss the physical system and present the Hamiltonian of the system. 
In this section different paired phases are also discussed
In Sec.~\ref{sec:nocor} we proceed to compute the noise correlation functions
for different paired phases at zero temperature and
in Sec.~\ref{sec:non_0_T} computations are generalized to non-zero temperatures.
In Sec.~\ref{sec:1D} noise correlations in a one dimensional lattice
are discussed while making a clear connection with the computed 1D phase-diagram.
We end with some concluding remarks in Sec.~\ref{sec:conc}.

\section{The Hamiltonian of the system}
\label{sec:sysham}
We assume in our calculation a three dimensional cubic optical lattice. 
The system is composed of a two component Fermi gas where components are  
different hyperfine states (for concreteness we assume  $^6\text{Li}$ atoms).
In terms of the field operators $\hat \Psi_{\sigma}({\bf r})$ the Hamiltonian of the system  is given by
\beq
\label{eq:ex_ham}
\begin{split}
&\hat H=\sum_{\sigma=\uparrow,\downarrow}\int \, d{\bf r}\, \hat \Psi_{\sigma}^\dagger({\bf r})\left(-\frac{\hbar^2 \nabla^2}{2m_{\sigma}}+V_{\sigma}({\bf r})\right)
\hat \Psi_{\sigma}({\bf r})+\\
&\int\int\,d{\bf r}\,d{\bf r'}\,  \hat \Psi_{\uparrow}^\dagger({\bf r}) \hat \Psi_{\downarrow}^\dagger({\bf r'})g({\bf r},{\bf r'})\hat \Psi_{\downarrow}({\bf r'})
\hat \Psi_{\uparrow}({\bf r})-\sum_{\sigma=\uparrow,\downarrow}\mu_{\sigma}\hat N_{\sigma}\\
&=\sum_{\sigma=\uparrow,\downarrow}\int \, d{\bf r}\, \hat \Psi_{\sigma}^\dagger({\bf r})\left(-\frac{\hbar^2 \nabla^2}{2m_{\sigma}}+V_{\sigma}({\bf r})-\mu_{\sigma}\right)
\hat \Psi_{\sigma}({\bf r})+\\
&\int\int\,d{\bf r}\,d{\bf r'}\,  \hat \Psi_{\uparrow}^\dagger({\bf r}) \hat \Psi_{\downarrow}^\dagger({\bf r'})g({\bf r},{\bf r'})\hat \Psi_{\downarrow}({\bf r'})
\hat \Psi_{\uparrow}({\bf r}),
\end{split}
\enq
where $\hbar=h/(2\pi)$ and $h$ is the Planck constant. $\mu_{\sigma}$ is
the chemical potential of the component $\sigma$ and the  lattice potential is given 
by $V_{\sigma}({\bf r})=sE_{r}\sum_{i=1}^3\sin^2(k x_j)$, where
$s$ is the lattice depth and  $E_r=\hbar^2 k^2/(2m)$ is the recoil energy ($k=\pi/d$ and $d$ is the lattice constant).
In the usual way the interaction between atoms is modeled by a contact potential  \[ g({\bf r},{\bf r'})=\frac{4\pi \hbar^2 a}{m}\delta({\bf r}-{\bf r'}),\]
where $a$ is the s-wave scattering length. The number operator  of the component $\sigma$ 
\[\hat N_{\sigma}=\int\, d{\bf r}\, \hat n_{\sigma}({\bf r})=\int\, d{\bf r}\, \hat \Psi_{\sigma}^{\dagger}({\bf r})\hat \Psi_{\sigma}({\bf r}),\]
is expressed in terms of the density operator $\hat n_{\sigma}({\bf r})= \hat \Psi_{\sigma}^{\dagger}({\bf r})\hat \Psi_{\sigma}({\bf r})$.
In our case the field-operators are fermionic which implies fermionic equal time anticommutation relations
\beq
\label{eq:anticom}
\begin{split}
&\{\hat \Psi_{\alpha}^{\dagger}({\bf r}),\hat \Psi_{\beta}({\bf r'})\}=\hat \Psi_{\alpha}^{\dagger}({\bf r})\hat \Psi_{\beta}({\bf r'})+
\hat \Psi_{\beta}({\bf r'})\hat \Psi_{\alpha}^{\dagger}({\bf r})=\delta_{\alpha\beta}\delta({\bf r}-{\bf r'})\\
&\{\hat \Psi_{\alpha}^{\dagger}({\bf r}),\hat \Psi_{\beta}^{\dagger}({\bf r'})\}=\{\hat \Psi_{\alpha}({\bf r}),\hat \Psi_{\beta}({\bf r'})\}=0,
\end{split}
\enq 
where $\delta_{\alpha\beta}$ is the Kronecker delta, and $\delta({\bf r}-{\bf r'})$ is the Dirac delta-function.

\subsection{The Hubbard model}

In a sufficiently 
deep lattice we can expand the field operators in terms of the well localized (lowest band) Wannier functions as
\beq
\label{eq:fieldop}
\begin{split}
\hat \Psi_{\sigma}({\bf r})=\sum_iw_i({\bf r})\hat c_{\sigma,i},
\end{split}
\enq
where $w_i({\bf r})$ is  Wannier function centered around a lattice point $i$, and $\hat c_{\sigma,i}$ 
is a fermionic annihilation operator which annihilate fermions of component $\sigma$ at site $i$.

In assuming that only the lowest band states are occupied we are
assuming temperatures
which are much lower than the bandgap. We can estimate the energy of the
vibrational levels
by approximating lattice wells with harmonic oscillators and in this way
find that temperatures
should satisfy a criterion
\beq
\label{eq:raja_lampo}
k_BT<<\frac{\hbar \pi}{d}\sqrt{\frac{2sE_r}{m}}
\enq
where $k_B$ is the Boltzmann constant.
If, for concreteness, we assume $^{6}{\rm Li}$ atoms in a cubic lattice
with
a lattice spacing $505\, {\rm nm}$, the above condition implies
temperatures much below
$5\mu\text{K}$ for a lattice depth of $10 {\rm E_r}$.

It is assumed that only tunneling  between  the nearest neighbors is of importance. This assumption of tight binding is reasonable when the Wannier functions are well 
localized. In other words,
overlap integrals between the next nearest neighbor Wannier functions are small compared to overlap integrals between  the nearest neighbor Wannier functions.
This always happens when the lattice 
is deep enough~\cite{Bloch2007a}.  Using this approximation one finds the Hubbard Hamiltonian~\cite{Jaksch1998a}
\beq
\label{eq:Hub_Ham}
\begin{split}
\hat H=&\sum_{ n}(-\mu_{\uparrow}\hat c_{\uparrow, n}^{\dagger} \hat c_{\uparrow, n}-
\mu_{\downarrow}\hat c_{\downarrow,n}^{\dagger} \hat c_{\downarrow,n})+
U\sum_{ n}\hat c_{\uparrow, n}^{\dagger}\hat c_{\downarrow,n}^{\dagger} \hat c_{\downarrow,n} \hat c_{\uparrow, n}\\
&-\left(J_x\sum_{\langle n, m\rangle_x}+J_y\sum_{\langle n,m\rangle_y}+J_z\sum_{\langle n, m\rangle_z}\right)
(\hat c_{\uparrow, m}^{\dagger} \hat c_{\uparrow, n}+\hat c_{\downarrow, n}^{\dagger} \hat c_{\downarrow, n}),
\end{split}
\end{equation}
where $\langle m,n \rangle_\alpha$ means a sum over the nearest neighbors in the $\alpha$-direction and hopping strength is defined by
\[J_{\alpha}=-\int\, d{{\bf r}}\, w_{n}^*({\bf r})\left(-\frac{\hbar^2 \nabla^2}{2m}+sE_r\sum_i\sin^2 (k x_i)\right)w_{n\pm d\hat x_{\alpha}}({\bf r}).\]
Furthermore, the dominant on site interaction coupling strength is given by
\[U=\frac{4\pi \hbar^2 a}{m}\int\, d{\bf r}\, |w_i({\bf r})|^4.\]
In our calculations we have approximated the Wannier functions with a harmonic oscillator ground state in each potential well. 

The most stringent criteria for temperatures follows from the fact that
critical temperature
for the superfluidity is typically much less than the Fermi energy,
which in turn is less
than the bandwidth. This means that in the regime of greatest interest,
the temperature should be
of same order or less than the tunneling energy scale $J$.

\subsection{The ground state ansatz at zero temperature}
On a mean-field level in the homogeneous system the ground state ansatz at zero temperature 
which includes the possibility of the BCS-state~\cite{Bardeen1957a}, the breach pair(BP)-state~\cite{Sarma1963a,Liu2003a}, 
and the plane wave FFLO-state~\cite{Fulde1964a,Larkin1964a} can be expressed as~\cite{Sheehy2007a}
\beq
\label{eq:GS_T_0}
|\Psi_{GS}\rangle=\prod_{{\bf k}\in G_3}\hat c_{\downarrow,-{\bf k}+{\bf q}}^{\dagger}\prod_{{\bf k}\in G_2}\hat c_{\uparrow,{\bf k}}^{\dagger}
\prod_{{\bf k}\in G_1}(u_{{\bf k},{\bf q}}+v_{{\bf k},{\bf q}}\hat c_{\uparrow,{\bf k}}^{\dagger}\hat c_{\downarrow,-{\bf k}+{\bf q}}^{\dagger})|0\rangle,
\enq
where $|0\rangle$ is the vacuum state.
In area $G_1$ both quasiparticle dispersions $E_{\uparrow,{\bf k},{\bf q}},E_{\downarrow,{\bf k},{\bf q}} $
are positive, in area
$G_2$ $E_{\downarrow,{\bf k},{\bf q}}\geq 0 $ and $E_{\uparrow,{\bf k},{\bf q}}< 0 $, and in area
$G_3$ $E_{\uparrow,{\bf k},{\bf q}}\geq 0 $ and $E_{\downarrow,{\bf k},{\bf q}}< 0 $. 
Area $G_1$ is populated  by pairs with a quasi-momentum ${\bf q}$.
Area $G_2$ is populated by  atoms of the component $\uparrow$ in the quasi-momentum states ${\bf k}\in G_2$   
while the area $G_3$ is populated by $\downarrow$-atoms 
in the quasi-momentum states $-{\bf k}+{\bf q}$. 
By diagonalizing the mean-field theory one finds the quasiparticle dispersions 
\beq
\label{eq:quaspar}
E_{\downarrow/\uparrow,{\bf k},{\bf q}}=
\sqrt{\left(\frac{\epsilon_{\uparrow,{\bf k}}+\epsilon_{\downarrow,-{\bf k}+{\bf q}}}{2}\right)^2+\Delta^2}\mp
\frac{\epsilon_{\uparrow,{\bf k}}-\epsilon_{\downarrow,-{\bf k}+{\bf q}}}{2},
\enq
where $\Delta$ is the pairing gap and the single particle dispersions are
\[\epsilon_{\sigma,{\bf k}}=\sum_{\alpha=1}^32J_\alpha(1-\cos (k_\alpha d))-\mu_{\sigma}.\]
The quasiparticle amplitudes $u_{{\bf k},{\bf q}}$ and $v_{{\bf k},{\bf q}}$ are given by
\beq
\begin{split}
&u_{{\bf k},{\bf q}}^2=\frac{1}{2}\left(1+
\frac{\epsilon_{\uparrow,{\bf k}}+\epsilon_{\downarrow,-{\bf k}+{\bf q}}}{2\sqrt{[(\epsilon_{\uparrow,{\bf k}}+\epsilon_{\downarrow,-{\bf k}+{\bf q}})/2]^2+\Delta^2}}\right)\\
&v_{{\bf k},{\bf q}}^2=1-u_{{\bf k},{\bf q}}^2.
\end{split}
\enq

Based on the properties of this ansatz, it can be classified further.
When the wavevector ${\bf q} \neq 0$, the ground state is called the FFLO-state which breaks the 
translational symmetry. If ${\bf q}=0$ and both 
quasiparticle dispersions $E_{\uparrow,{\bf k}},E_{\downarrow,{\bf k}}$  are positive for all ${\bf k}$-vectors, 
the ground state is called BCS-state, and finally if 
${\bf q}=0$ and $E_{\sigma,{\bf k}}<0$ for some ${\bf k}$-vectors, the ground state is called the BP-state. 
This last case involves phase separation in the momentum space.

\begin{figure}
\includegraphics[scale=0.6]{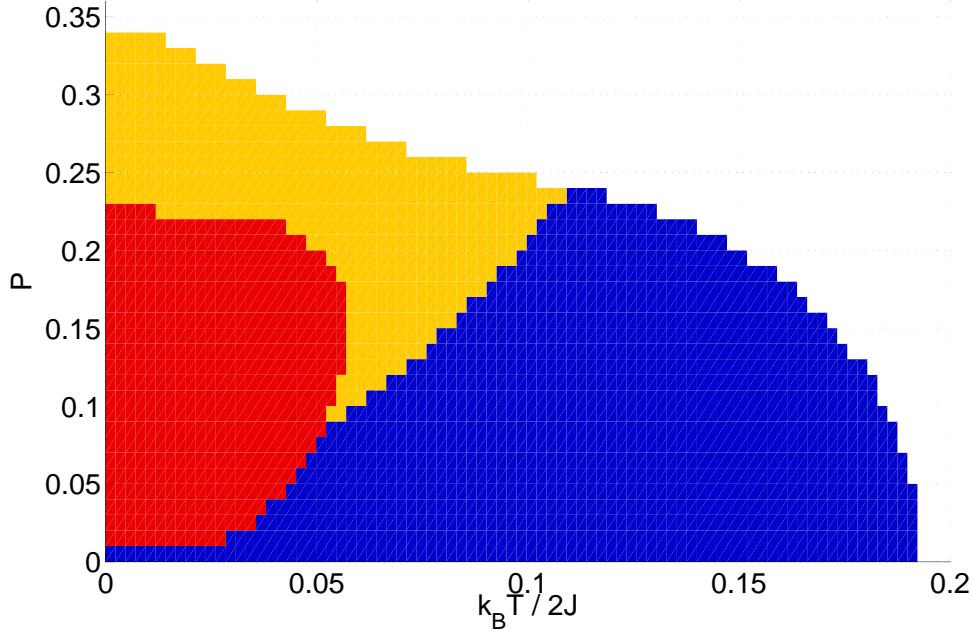}
\caption{[Color online]  The phase diagram of Fermi gas in a 3D-lattice~\cite{Koponen2007a}.
Colors (or shading): BCS/BP states=blue or dark gray, FFLO=yellow or light gray, phase separation=red or gray, and normal gas=white.
The parameters where such that 
the average filling fraction $(n_{\uparrow}+n_{\downarrow})/2=0.2$, $J=0.07E_r$, and $U/(2J)=-1.86$.}
\label{Fig:3Dphasediagram}
\end{figure}

In Fig.~\ref{Fig:3Dphasediagram} we show on example phase diagram of the
polarized
system in a three-dimensional lattice~\cite{Koponen2007a}. At low
temperatures,
phase separation is favored for low polarization, while for larger
polarization
FFLO state is energetically favorable in substantially large parts of
the phasediagram.
At non-zero temperatures BP-state
can be energetically stable and occupy regions of the phasediagram close
to
the critical temperature. Although we do not show it in the figure,
for low polarization quasi-particle dispersions
of the BP-state can be gapped while for higher polarization they are
gapless. Very close to the critical temperature the dispersions are always
gapless for the polarized system.
Note that only BCS-normal phase separation was considered, not BCS-FFLO coexistence 
which could make the parameter window where FFLO occurs even wider.

\section{Noise correlations}
\label{sec:nocor}
Our aim is to compute the noise correlation functions which are defined by~\cite{Altman2004a}
\beq
\label{eq:kormaar}
\begin{split}
&G_{\alpha \beta}({\bf r},{\bf r'})=\langle \hat n_{\alpha}({\bf r},t)\hat n_{\beta}({\bf r'},t)\rangle-
\langle \hat n_{\alpha}({\bf r},t)\rangle\langle \hat n_{\beta}({\bf r'},t)\rangle\\
&=\langle \hat \Psi_{\alpha}^{\dagger}({\bf r},t) \hat \Psi_{\beta}^{\dagger}({\bf r'},t)\hat \Psi_{\beta}({\bf r'},t) \hat \Psi_{\alpha}({\bf r},t)\rangle
+\delta_{\alpha\beta}\delta({\bf r}-{\bf r'})\langle \hat n_{\alpha}({\bf r},t)\rangle-\langle \hat n_{\alpha}({\bf r},t)\rangle\langle \hat n_{\beta}({\bf r'},t)\rangle,
\end{split}
\enq
where $\alpha$ and $\beta$ are component indeces, ${\bf r}$ and ${\bf r'}$ are the positions, 
$t$ is time. The term with Dirac's delta function is due to the normal ordering of the density-density
expectation value.

Often in experiments a gas which has been trapped is released and the gas expands.
When collisions during the expansion can be ignored and the expansion is ballistic,
the density-density correlations after the expansion reflect
correlations in momentum space at $t = 0$. Therefore, correlations can be computed using the wave-
function in momentum space prior to expansion. Positions in real space after time
$t$ are related to k-vectors trough ${\bf r} = t{\bf k}/m$. 
In other words at long times (the saddle point approximation of the free evolution amounts to the limit $t\gg m d^2/\hbar$) \[w_i({\bf r},t)\sim e^{-iQ({\bf r})\cdot {\bf R}_i},\]
where ${\bf R}_i=d(i_x,i_y,i_z)$ ($i_n$ are integers) is a lattice vector and \[Q({\bf r})=\frac{m{\bf r}}{\hbar t}.\]
Thus the densities become
\beq
\label{eq:den}
n_{\alpha}({\bf r},t)=\langle \hat n_{\alpha}({\bf r},t) \rangle=A(t)^2\sum_{i,j}e^{iQ({\bf r})\cdot {\bf R}_{ij}}\langle \hat c_{\alpha,i}^{\dagger}\hat c_{\alpha,j}\rangle,
\enq
where ${\bf R}_{ij}={\bf R}_i-{ \bf R}_j$ and $A(t)$ is a time dependent scale factor which depends on the Wannier functions. 
The correlation functions in Eq.~\eqref{eq:kormaar}  now become
\beq
\label{eq:kormaar2}
\begin{split}
&G_{\alpha\beta}({\bf r},{\bf r'})=A(t)^4\sum_{i,j,m,n}e^{iQ({\bf r})\cdot {\bf R}_{ij}+iQ({\bf r'})\cdot {\bf R}_{mn}}\langle \hat c_{\alpha,i}^{\dagger}
\hat c_{\beta,m}^{\dagger}\hat c_{\beta,n}\hat c_{\alpha,j}\rangle\\
&+\delta_{\alpha\beta}\delta({\bf r}-{\bf r'})\langle \hat n_{\alpha}({\bf r},t)\rangle-\langle \hat n_{\alpha}({\bf r},t)\rangle\langle \hat n_{\beta}({\bf r'},t)\rangle.
\end{split}
\enq

\subsection{The noise correlations of the BCS-state at zero temperature}
\label{subsec:BCS_0}
We review here the noise correlations of the BCS-state.
The usual unpolarized BCS ground state is extracted from the ansatz in
Eq.~\eqref{eq:GS_T_0}
by setting both chemical potentials to be equal and ${\bf q=0}$. For the BCS-state
both quasiparticle dispersions are positive.
Because the BCS-ground state  is presented in momentum space, for simplicity we express the densities in Eq.~\eqref{eq:den}
in terms of the operators in the momentum space 
\beq
\label{eq:fourierdens}
\langle \hat n_{\alpha}({\bf r},t)\rangle=\frac{A(t)^2}{M^2}\sum_{i,j,{\bf k}_1,{\bf k}_2}e^{iQ({\bf r})\cdot {\bf R}_{ij}}e^{-i{\bf k}_1\cdot{\bf R}_i+i{\bf k}_2\cdot{\bf R}_j}
\langle \hat c_{\alpha,{\bf k}_1}^{\dagger}\hat c_{\alpha,{\bf k}_2}\rangle
\enq
 and similarly for the noise correlations in Eq.~\eqref{eq:kormaar2}
\beq
\label{eq:fouriercor}
\begin{split}
G_{\alpha\beta}({\bf r},{\bf r'})=&\frac{A(t)^4}{M^4}\sum_{\substack {i,j,m,n,\\{\bf k}_1,{\bf k}_2,{\bf k}_3,{\bf k}_4} }\bigg(e^{iQ({\bf r})\cdot {\bf R}_{ij}+iQ({\bf r'})\cdot {\bf R}_{mn}}
e^{-i{\bf k}_1\cdot{\bf R}_i+i{\bf k}_2\cdot{\bf R}_j-i{\bf k}_3\cdot{\bf R}_m+i{\bf k}_4\cdot{\bf R}_n}\\
&\langle \hat c_{\alpha,{\bf k}_1}^{\dagger}\hat c_{\beta,{\bf k}_3}^{\dagger}\hat c_{\beta,{\bf k}_4}\hat c_{\alpha,{\bf k}_2}\rangle\bigg)\\
&+\delta_{\alpha\beta}\delta({\bf r}-{\bf r'})\langle \hat n_{\alpha}({\bf r},t)\rangle-\langle\hat n_{\alpha}({\bf r},t)\rangle\langle\hat n_{\beta}({\bf r'},t)\rangle,
\end{split}
\enq
where $M$ is the number of lattice sites. Now one needs the expectation values 
$\langle \hat c_{\alpha,{\bf k}_1}^{\dagger}\hat c_{\beta,{\bf k}_3}^{\dagger}\hat c_{\beta,{\bf k}_4}\hat c_{\alpha,{\bf k}_2}\rangle$ and 
$\langle \hat c_{\alpha,{\bf k}_1}^{\dagger}\hat c_{\alpha,{\bf k}_2}\rangle$.
For the latter term one gets
\beq
\label{eq:densexval}
\langle \Psi_{BCS}|\hat c_{\alpha,{\bf k}_1}^{\dagger}\hat c_{\alpha,{\bf k}_2}|\Psi_{BCS}\rangle=\delta_{{\bf k}_1{\bf k}_2}|v_{{\bf k}_1}|^2 
\enq
and using this the densities in Eq.~\eqref{eq:fourierdens} can be expressed as
\beq
\label{eq:BCS-dens}
\langle \hat n_{\alpha}({\bf r},t)\rangle=\frac{A(t)^2}{M^2}\sum_{i,j,{\bf k}}e^{i(Q({\bf r})-{\bf k})\cdot {\bf R}_{ij}}|v_{\bf k}|^2.
\enq 
Now the number of particles is the same for both the components as it should be.
When the lattice is large, i.e. when the lattice size $M\gg 1$, the exponential terms only 
add constructively when $Q({\bf r})-{\bf k}=0$ and we can 
approximate the densities
as 
\beq
\label{eq:BCS-dens_delta}
\langle \hat n_{\alpha}({\bf r},t)\rangle=A(t)^2\sum_{\bf k}\delta\left({\bf r}-\frac{\hbar t}{m}{\bf k}\right)|v_{\bf k}|^2.
\enq 

Non-vanishing  four operator expectation values are
\beq
\label{eq:fourexp}
\begin{split}
&\langle \Psi_{BCS}|\hat c_{\uparrow,{\bf k}}^{\dagger}\hat c_{\downarrow,-{\bf k'}}^{\dagger}\hat c_{\downarrow,-{\bf k'}}\hat c_{\uparrow,{\bf k}}|\Psi_{BCS}\rangle=
\delta_{{\bf k}{\bf k'}}|v_{\bf k}|^2+(1-\delta_{{\bf k}{\bf k'}})|v_{\bf k'}|^2|v_{\bf k}|^2\\
&\langle \Psi_{BCS}|\hat c_{\uparrow (\downarrow),{\bf k}}^{\dagger}\hat c_{\uparrow (\downarrow),{\bf k'}}^{\dagger}
\hat c_{\uparrow (\downarrow),{\bf k'}}\hat c_{\uparrow (\downarrow),{\bf k}}|\Psi_{BCS}\rangle=(1-\delta_{{\bf k}{\bf k'}})|v_{\bf k'}|^2|v_{\bf k}|^2
\end{split}
\enq
and combining these results and taking the limit of large lattices, we find
\beq
\label{eq:BCScorrelations2}
\begin{split}
&G_{\uparrow\downarrow}({\bf r},{\bf r'})=G_{\downarrow\uparrow}({\bf r},{\bf r'})=A(t)\sum_{{\bf k}}|u_{\bf k}|^2|v_{\bf k}|^2
\delta\left({\bf r}-\frac{\hbar t\tilde{\bf k}}{m}\right)\delta\left({\bf r'}+\frac{\hbar t\tilde{\bf k'}}{m}\right)\\
&G_{\sigma\sigma}({\bf r},{\bf r'})=-A(t)^4\sum_{{\bf k}}|v_{\bf k}|^4
\delta\left({\bf r}-\frac{\hbar t\tilde{\bf k}}{m}\right)\delta\left({\bf r'}-\frac{\hbar t\tilde{\bf k'}}{m}\right)\\
&+\delta({\bf r}-{\bf r'})\langle \hat n_{\sigma}({\bf r},t)\rangle,
\end{split}
\enq
where $\tilde{ \bf k}={ \bf k}+\sum_{\alpha=1}^32n_\alpha\pi\hat x_\alpha/d $ and $\tilde{ \bf k'}={ \bf k}+\sum_{\alpha=1}^32m_\alpha\pi\hat x_\alpha/d$,
where ${ \bf k}$ is a lattice momentum, $\hat x_\alpha$ are orthogonal unit vectors and $n_i$ and $m_i$ are integers.
Because $|u_{\bf k}|^2|v_{\bf k}|^2\sim \Delta^2$
the noise correlation $G_{\uparrow\downarrow}$
between different components vanishes with vanishing gap $\Delta$ and is thus identically zero in the normal state.
One can see that the noise correlations in the BCS-state  differ crucially from this.
Because in the BCS-state momenta ${\bf k}$ and $-{\bf k}$ are correlated, the noise correlation signal is pronounced only 
if \[{\bf r}+{\bf r'}=\sum_{i=1}^3\frac{\hbar 2n_i\pi t\hat x_i}{md}. \]
The continuum result~\cite{Altman2004a} is similar, 
but peaks appear when ${\bf r}+{\bf r'}=0$. In the continuum limit our result reduces to the known result.

On the other hand correlations of the single component $G_{\uparrow\uparrow}$ and $G_{\downarrow\downarrow}$ 
show a hole when \[{\bf r}-{\bf r'}=\sum_{i=1}^3\frac{2\hbar n_i\pi t\hat x_i}{md}\] and $n_i\neq 0$
for at least one $i$~\cite{Andersen2007a}.
The reason for this is the fermionic antibunching related to 
Pauli blocking i.e. two or more identically
fermions cannot be in the same momentum-state. This fermionic antibunching was also measured in the noise correlations 
of an ideal Fermi gas~\cite{Rom2006a}.

\begin{figure}
\begin{tabular}{lll}
\includegraphics{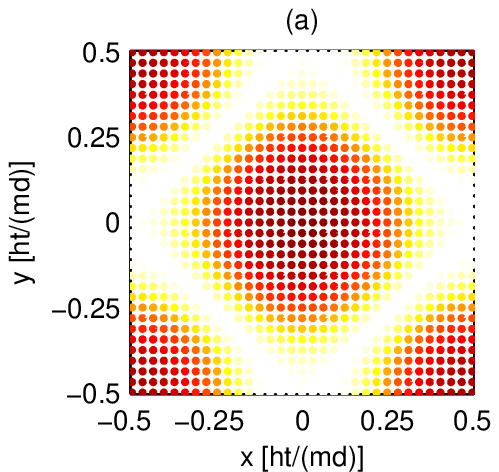} & \includegraphics{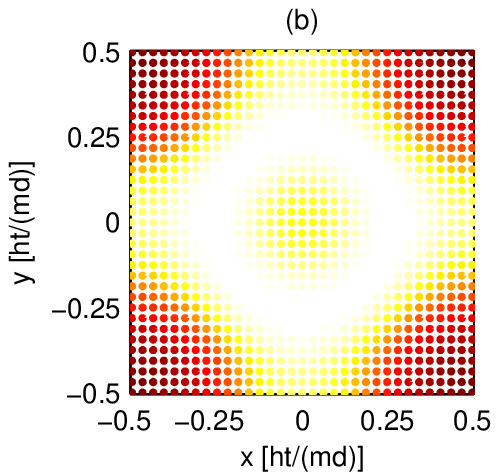} & \includegraphics{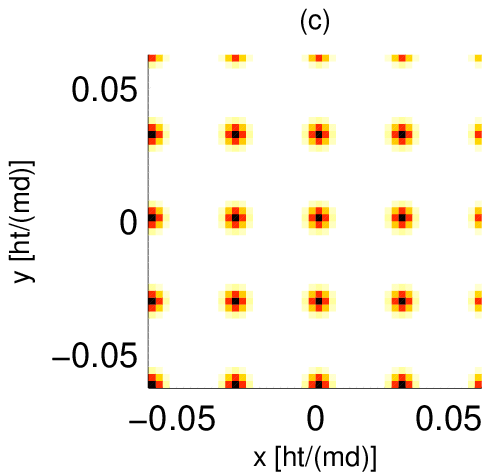}
\end{tabular}
\caption{[Color online] In (a) we show a cut of 
the BCS noise correlations between components in the $z=0$ plane. 
In (b) we show the integrated  BCS correlations between the components. 
In (c) we demonstrate the noise correlation of a single component in the plane $z=0$.
As parameters we used the average filling fraction 
$(n_{\uparrow}+n_{\downarrow})/2=0.30$, polarization $P=(n_{\uparrow}-n_{\downarrow})/(n_{\uparrow}+n_{\downarrow})=0.0$,
$\Delta/(2J)=1.03$ and $U/(2J)=-3.0$. All these examples are calculated at $T=0$. 
In (a) and (b) we choose ${\bf r'}=-{\bf r}$ and in (c)
we choose  $y'=y$, $z=z'=0$, and $x'=x+\hbar t 2\pi/(md)$. Color-coding is such that light colors
imply high peaks and dark colors low. }
\label{fig:fig2}
\end{figure}

In Fig.~\ref{fig:fig2} we demonstrate the BCS noise correlations at
$T=0$. In Fig.~\ref{fig:fig2} (a) we show a cut in the $z=0$ plane of
the BCS noise correlation between the components while 
in Fig.~\ref{fig:fig2} (b) we plot  the column integrated BCS-state noise correlations 
\[G_{\uparrow,\downarrow}(x,y)=\int\, dz\, G_{\uparrow,\downarrow}\left(x,y,z,-x+\hbar tq_x/m,-y+\hbar tq_y/m,-z+\hbar tq_z/m\right).\]
As one can see from the Fig.~\ref{fig:fig2}, the noise correlations reflect underlying lattice structure
and strongly non-spherical structure of the Fermi surface.
We can see also from Fig.~\ref{fig:fig2} (a) and (b) that the peaks are highest near the Fermi surfaces
where $\epsilon_{\uparrow,{\bf k}}+\epsilon_{\downarrow,-{\bf k}}=0$.
In the weakly interacting BCS limit correlations are strongly peaked
at the Fermi surface, but the distribution of the peak heights becomes broad in the BEC
limit, and the signal approaches the interference signal of the Bose-Einstein condensate.

In Fig.~\ref{fig:fig2} (c) we show the BCS state correlations  
of a single component in the $z=0$ plane. As is clear, the BCS-state noise correlation
of a single component shows  similar antibunching behavior as the ideal Fermi gas noise correlation. 
If ${\bf r}={\bf r'}+\sum_{i=1}^32\hbar n_i\pi t\hat x_i/(md)$ and  and $n_i\neq 0$
for at least one $i$, then the noise correlation of a single component shows the holes.
However, the BCS-state noise correlation
of a single component does also show the bunching peak when ${\bf r}={\bf r'}$.
This result differs from the ideal gas result where such a peak is absent.

\subsection{The noise correlations of the BP-state at zero temperature}
\label{subsec:BP_0}
The ground state in Eq.~\eqref{eq:GS_T_0} is the BP-state when ${\bf q}=0$ and one of the quasiparticle dispersions 
$E_{\sigma,{\bf k}}$ is negative 
for some values of the momentum ${\bf k}$.
This implies that in the BP-state $|\delta\mu|=|\mu_{\uparrow}-\mu_{\downarrow}|>2\Delta$. 
The BP-state is a gapless state, because one of the quasiparticle dispersions
changes its sign when the momentum varies. The BP state at zero-temperature is expressed as
\beq
\label{eq:BPgs}
|\Psi_{BP}\rangle=\prod_{{\bf k}\in G_3}\hat c_{\downarrow,-{\bf k}}^{\dagger}\prod_{{\bf k}\in G_2}\hat c_{\uparrow,{\bf k}}^{\dagger}
\prod_{{\bf k}\in G_1}(u_{{\bf k}}+v_{{\bf k}}\hat c_{\uparrow,{\bf k}}^{\dagger}\hat c_{\downarrow,-{\bf k}}^{\dagger})|0\rangle.
\enq
In the area $G_1$ both quasiparticle dispersions $E_{\uparrow,{\bf k},{\bf q}},E_{\downarrow,{\bf k},{\bf q}}\geq 0 $, in the area
$G_2$ $E_{\downarrow,{\bf k}}\geq 0 $ and $E_{\uparrow,{\bf k}}< 0 $, and in the area
$G_3$ $E_{\uparrow,{\bf k}}\geq 0 $ and $E_{\downarrow,{\bf k}}< 0 $.

Now we can calculate the required expectation values similarly as before. 
We find that the densities in the BP-state (in a large lattice) are 
\beq
\label{eq:evBPdens3}
\begin{split}
\langle \hat n_{\uparrow (\downarrow)}({\bf r},t)\rangle=A(t)^2\sum_{\bf k}\delta\left({\bf r}-\frac{\hbar t {\bf k}}{m}\right)
[(1-\theta(E_{\uparrow (\downarrow),{\bf k}})+\theta(E_{\uparrow,{\bf k}})\theta(E_{\downarrow,{\bf k}})|v_{\bf k}|^2]
\end{split}
\enq
and non-vanishing expectation values in the noise correlations in Eq.~\eqref{eq:fouriercor} are in turn given by
\beq
\label{eq:evBPcor}
\begin{split}
&\langle\Psi_{BP}|\hat c_{\uparrow,{\bf k}}^{\dagger}\hat c_{\downarrow,-{\bf k'}}^{\dagger}\hat c_{\downarrow,-{\bf k'}}\hat c_{\uparrow,{\bf k}}|\Psi_{BP}\rangle=\\
&(1-\delta_{\bf kk'})[(1-\theta(E_{\uparrow,{\bf k}}))(1-\theta(E_{\downarrow,{\bf k'}}))+
\theta(E_{\uparrow,{\bf k}})\theta(E_{\downarrow,{\bf k}})\theta(E_{\uparrow,{\bf k'}})\theta(E_{\downarrow,{\bf k'}})|v_{\bf k}|^2|v_{\bf k'}|^2\\
&+\theta(E_{\uparrow,{\bf k}})\theta(E_{\downarrow,{\bf k}})(1-\theta(E_{\downarrow,{\bf k'}}))|v_{\bf k}|^2
+\theta(E_{\uparrow,{\bf k'}})\theta(E_{\downarrow,{\bf k'}})(1-\theta(E_{\uparrow,{\bf k}}))|v_{\bf k'}|^2]\\
&+\delta_{\bf kk'}\theta(E_{\uparrow,{\bf k}})\theta(E_{\downarrow,{\bf k}})|v_{\bf k}|^2\\
&\langle\Psi_{BP}|\hat c_{\uparrow (\downarrow),{\bf k}}^{\dagger}\hat c_{\uparrow (\downarrow),{\bf q}}^{\dagger}
\hat c_{\uparrow (\downarrow),{\bf k'}}\hat c_{\uparrow (\downarrow),{\bf k}}|\Psi_{BP}\rangle=\\
&(1-\delta_{\bf kk'})[(1-\theta(E_{\uparrow (\downarrow),{\bf k}}))(1-\theta(E_{\uparrow (\downarrow) ,{\bf k'}}))+
\theta(E_{\uparrow,{\bf k}})\theta(E_{\downarrow,{\bf k}})\theta(E_{\uparrow,{\bf q}})\theta(E_{\downarrow,{\bf k'}})|v_{\bf k}|^2|v_{\bf k'}|^2\\
&+\theta(E_{\uparrow,{\bf k}})\theta(E_{\downarrow,{\bf k}})(1-\theta(E_{\uparrow (\downarrow),{\bf k'}}))|v_{\bf k}|^2
+\theta(E_{\uparrow,{\bf k'}})\theta(E_{\downarrow,{\bf k'}})(1-\theta(E_{\uparrow, (\downarrow){\bf k}}))|v_{\bf k'}|^2]].
\end{split}
\enq
Subtracting the densities cancels out most of these terms
and again for large lattices we get strongly peaked behaviour and
\beq
\label{eq:BPdeltacorr}
\begin{split}
&G_{\uparrow\downarrow(\downarrow\uparrow)}({\bf r},{\bf r'})=A(t)^4\sum_{{\bf k}}\delta\left({\bf r}-\frac{\hbar t \tilde{\bf k}}{m}\right)
\delta\left({\bf r'}+\frac{\hbar t \tilde{\bf k'}}{m}\right)
\theta(E_{\uparrow,{\bf k}})\theta(E_{\downarrow,{\bf k}})|u_{\bf k}|^2|v_{\bf k}|^2\\
&G_{\sigma\sigma}({\bf r},{\bf r'})=-A(t)^4\sum_{{\bf k}}\bigg[
\delta\left({\bf r}-\frac{\hbar t\tilde{\bf k} }{m}\right)\delta\left({\bf r'}-
\frac{\hbar t\tilde{\bf k'} }{m}\right)\theta(E_{\uparrow,{\bf k}})\theta(E_{\downarrow,{\bf k}})|v_{\bf k}|^4\\
&+(1-\theta(E_{\sigma,{\bf k}}))\bigg]
+\delta({\bf r}-{\bf r'})\langle \hat n_{\sigma}({\bf r},t)\rangle.
\end{split}
\enq
In the correlation between different components there are now areas where peaks are absent. 
The reason for this is that the BP-state has
phase separation in the momentum space. Because in the BP-state there is no pairing  between momenta ${\bf k}$ and $-{\bf k}$, if  
$E_{\uparrow,{\bf k}}<0$ or  $ E_{\downarrow,{\bf k}}<0$, in contrast to the BCS-state, there are no correlations between the points 
\[{\bf r}+{\bf r'}=\sum_{i=1}^3\frac{\hbar 2n_i\pi t\hat x_i}{md},\] in this region.

\begin{figure}
\begin{tabular}{ll}
\includegraphics{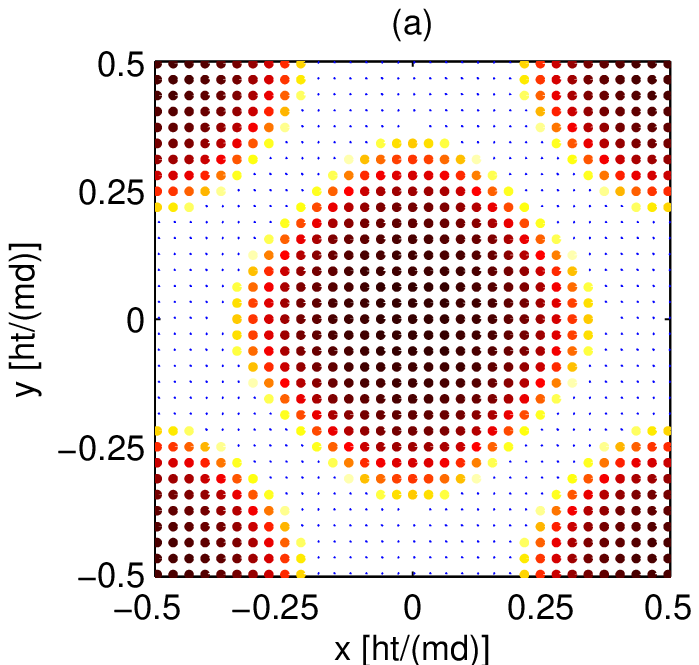} & \includegraphics{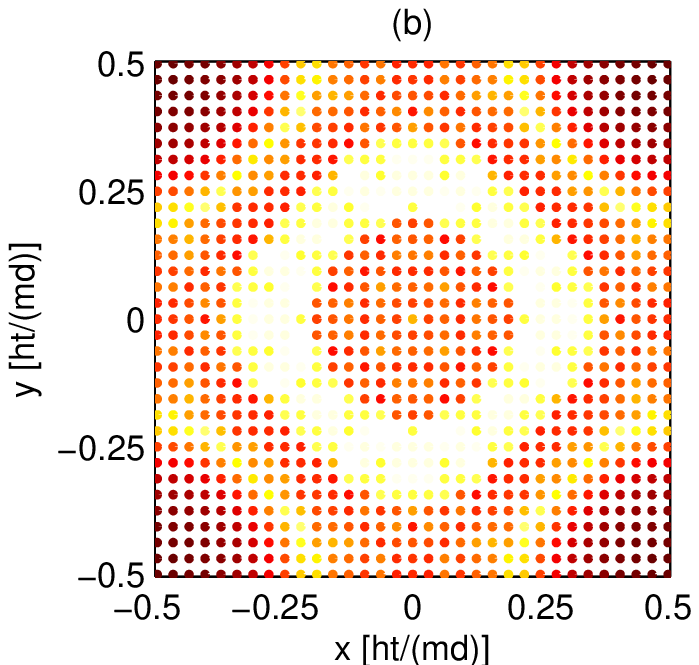}
\end{tabular}
\caption{[Color online] In this figure  we demonstrate examples of the
  BP noise correlations between the components at $T=0$. 
In (a) we demonstrate a cut in the $z=0$ plane while in (b) we show 
the column integrated BP noise correlation.
Parameters we used were $\Delta/(2J)=0.318$, $(\mu_{\uparrow}+\mu_{\downarrow })/2.0=2.18$, 
and $\delta\mu=\mu_{\uparrow}-\mu_{\downarrow }=1.14$.
In figure (a) the correlations vanish in the dotted area.
We have chosen ${\bf r'}=-{\bf r}$ and  color-coding is such that light colors 
imply high peaks and dark colors low.}
\label{fig:fig3}
\end{figure}

In Fig.~\ref{fig:fig3} we demonstrate the BP-state noise correlations at $T=0$.
In  Fig.~\ref{fig:fig3} (a) we have plotted the BP-state noise
correlation at the $z=0$ plane
while in Fig.~\ref{fig:fig3} (b) we show the corresponding column integrated noise correlation.
From Fig.~\ref{fig:fig3} (a) one can see clear areas without correlation peaks in the BP noise correlation due to the fact that
these areas are populated only by atoms of the majority component which are unpaired.   
From  the integrated correlations in Fig.~\ref{fig:fig3} (b)  we  can see that the peaks are now relatively low
in the area between the Fermi surfaces. This differs substantially from the integrated BCS noise correlation (see Fig.~\ref{fig:fig2} (b)).

Results in this section are intended mainly as 
a reference since the BP-state is not expected to be the ground state at zero temperature. 
Depending on polarization the 
ground state at zero temperature can be the BCS-state, phase separation~\cite{Bedaque2003a}, or the FFLO type state~\cite{Koponen2007a}.
However, at non-zero temperatures the BP-state can be the minimum of 
the free energy~\cite{Martikainen2006a}.

\subsection{The noise correlations of the plane wave FFLO-state at zero temperature}
\label{subsec:FFLO_0}
The FFLO ground state is presented in Eq.~\eqref{eq:GS_T_0}. Contrary to the BCS-state and the BP-state, in the FFLO-state  the parameter ${\bf q}\neq 0$
and the translational invariance is broken. In this case the order parameter is given by $\Delta({\bf R}_i)=\Delta_0\exp(i{\bf q}\cdot{\bf R}_i)$,
where ${\bf R}_i$ is a lattice vector.
In the FFLO-state the other  Fermi surface has been effectively shifted by the wavevector ${\bf q}$.  

The densities in the FFLO-state are (in the limit of large lattices) given by
\beq
\label{eq:evFFLOdens2}
\begin{split}
&\langle \hat n_{\uparrow}({\bf r},t)\rangle=A(t)^2\sum_{{\bf k}}\delta\left({\bf r}-\frac{\hbar t {\bf k}}{m}\right)
[(1-\theta(E_{\uparrow,{\bf k},{\bf q}})+\theta(E_{\uparrow,{\bf k},{\bf q}})\theta(E_{\downarrow,{\bf k},{\bf q}})|v_{{\bf k},{\bf q}}|^2]\\
&\langle \hat n_{\downarrow}({\bf r},t)\rangle=A(t)^2\sum_{{\bf k}}\delta\left({\bf r}-\frac{\hbar t ({\bf q}-{\bf k})}{m}\right)
[(1-\theta(E_{\downarrow,{\bf k},{\bf q}})+\theta(E_{\uparrow,{\bf k},{\bf q}})\theta(E_{\downarrow,{\bf k},{\bf q}})|v_{{\bf k},{\bf q}}|^2].
\end{split}
\enq
In the same way that we computed the noise correlations in the BCS- and BP-states, we find 
the noise correlation in the FFLO-state as
\beq
\label{eq:FFLOdeltacorr}
\begin{split}
&G_{\uparrow\downarrow}({\bf r},{\bf r'})=A(t)^4\sum_{{\bf k}}\delta\left({\bf r}-\frac{\hbar t \tilde{\bf k}}{m}\right)
\delta\left({\bf r'}+\frac{\hbar t(\tilde{\bf k'}-{\bf q})}{m}\right)
\theta(E_{\uparrow,{\bf k},{\bf q}})\theta(E_{\downarrow,{\bf k},{\bf q}})|u_{{\bf k},{\bf q}}|^2|v_{{\bf k},{\bf q}}|^2\\
&G_{\downarrow\uparrow}({\bf r},{\bf r'})=A(t)^4\sum_{{\bf k}}\delta\left({\bf r}+\frac{\hbar t (\tilde{\bf k}-{\bf q}) }{m}\right)
\delta\left({\bf r'}-\frac{\hbar t \tilde{\bf k'}}{m}\right)
\theta(E_{\uparrow,{\bf k},{\bf q}})\theta(E_{\downarrow,{\bf k},{\bf q}})|u_{{\bf k},{\bf q}}|^2|v_{{\bf k},{\bf q}}|^2\\
&G_{\uparrow\uparrow}({\bf r},{\bf r'})=-A(t)^4\sum_{{\bf k}}\bigg[
\delta\left({\bf r}-\frac{\hbar t\tilde{\bf k} }{m}\right)\delta\left({\bf r'}-\frac{\hbar t\tilde{\bf k'} }{m}\right)\theta(E_{\uparrow,{\bf k},{\bf q}})
\theta(E_{\downarrow,{\bf k},{\bf q}})|v_{{\bf k},{\bf q}}|^4\\
&+(1-\theta(E_{\uparrow ,{\bf k},{\bf q}}))\bigg]
+\delta({\bf r}-{\bf r'})\langle \hat n_{\uparrow}({\bf r},t)\rangle\\
&G_{\downarrow\downarrow}({\bf r},{\bf r'})=-A(t)^4\sum_{{\bf k}}\bigg[
\delta\left({\bf r}-\frac{\hbar t (\tilde{\bf k}-{\bf q}) }{m}\right)\delta\left({\bf r'}-\frac{\hbar t(\tilde{\bf k'}-{\bf q}) }{m}\right)\cdot \\
&\theta(E_{\uparrow,{\bf k},{\bf q}})\theta(E_{\downarrow,{\bf k},{\bf q}})|v_{{\bf k},{\bf q}}|^4+(1-\theta(E_{\downarrow ,{\bf k},{\bf q}}))\bigg]
+\delta({\bf r}-{\bf r'})\langle \hat n_{\downarrow}({\bf r},t)\rangle.
\end{split}
\enq
Because in the FFLO-state momenta ${\bf k}$ and $-{\bf k}+{\bf q}$ are
correlated, the points 
\[{\bf r}+{\bf r'}+\frac{\hbar t}{m}{\bf q}=\sum_{i=1}^3\frac{\hbar 2n_i\pi t\hat x_i}{md}\]
are now correlated in the expanded cloud. Since  the FFLO-state arises only when there is some polarization in the system,
at $T=0$ the FFLO-state is always a gapless state 
i.e. one of the quasiparticle dispersions $E_{\sigma,{\bf k},{\bf q}}$
changes its sign when momentum ${\bf k}$ varies.

Multimode FFLO-states also leave clear signatures on the noise correlations.
For example, for the two mode FFLO-state i.e. the state 
where the gap is given by $\Delta({\bf{R}_i})=\Delta_0\cos ({\bf q}\cdot{\bf R}_i)$, 
the order parameter can be written as
\[\begin{split} 
&\Delta({\bf R}_i)=\Delta_0\cos ({\bf q}\cdot{\bf R}_i)=\Delta_0\left(\frac{e^{i{\bf q}\cdot{\bf R}_i}+e^{-i{\bf q}\cdot{\bf R}_i}}{2}\right)\\
&=-U\langle \hat c_{\uparrow,i}\hat c_{\downarrow,i}\rangle=
\frac{-U}{M}\sum_{{\bf k},{\bf k'}}e^{-i({\bf k}+{\bf k'})\cdot{\bf R}_i}\langle \hat c_{\uparrow,{\bf k}}\hat c_{\downarrow,{\bf k'}}\rangle\\
&=e^{i{\bf q}\cdot{\bf R}_i}\frac{-U}{M}\sum_{\bf k}\langle \hat c_{\uparrow,{\bf k}}\hat c_{\downarrow,-{\bf k}-{\bf q}}\rangle+e^{-i{\bf q}\cdot{\bf R}_i}
\frac{-U}{M}\sum_{\bf k}\langle \hat c_{\uparrow,{\bf k}}
\hat c_{\downarrow,-{\bf k}+{\bf q}}\rangle.
\end{split}
\]
Therefore the only non-vanishing expectation values in the momentum space are 
$\langle\hat c_{\uparrow,{\bf k}}\hat c_{\downarrow,-{\bf k}+{\bf q}}\rangle$ 
and $\langle\hat c_{\uparrow,{\bf k}} \hat c_{\downarrow,-{\bf k}-{\bf q}}\rangle$. 
This implies that an $\uparrow$-atom in the momentum state ${\bf k}$ is paired with $\downarrow$-atoms 
in the momentum states $-{\bf k}+{\bf q}$ and  $-{\bf k}-{\bf q}$. Because in the single
mode FFLO-state points ${\bf r}$ and $-{\bf r}+\hbar t{\bf q}/m$ are correlated after free expansion,
in the two mode FFLO-state ${\bf r}$ is correlated with $-{\bf r}+\hbar t{\bf q}/m$ 
and $-{\bf r}-\hbar t{\bf q}/m$ after free expansion. For this reason with the two mode FFLO-state one can see 
pronounced correlation peaks, when ${\bf r}+{\bf r'}\pm\hbar t{\bf q}/m=0$.
Actually for the multimode FFLO-state, one could use noise correlations to perform a "Fourier analysis'' of 
the periodic order parameter and probe more complicated spatial dependencies than those discussed here.

\begin{figure}
\begin{tabular}{ll}
\includegraphics{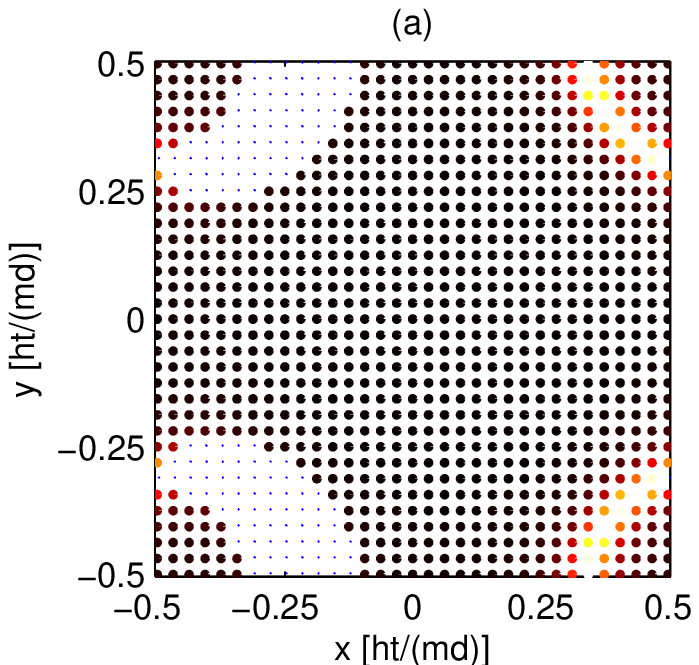} & \includegraphics{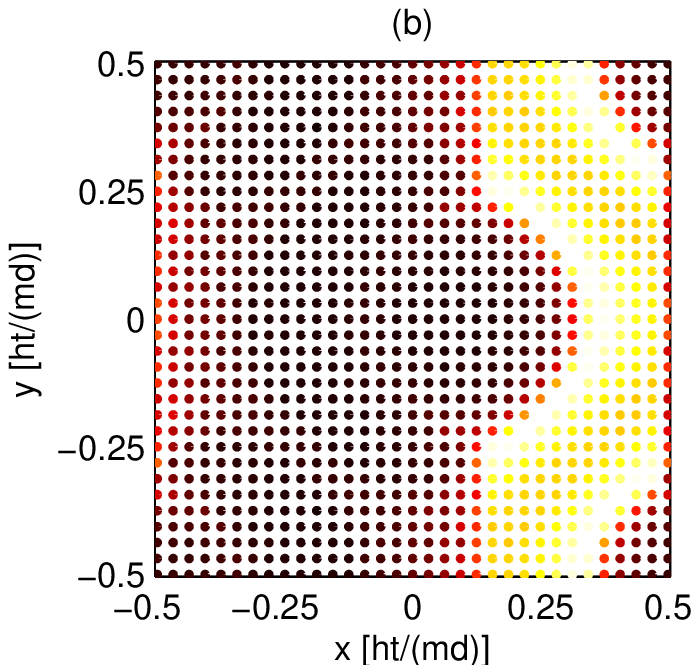}
\end{tabular}
\caption{[Color online] In this figure we show the FFLO-state noise correlation ($G_{\uparrow\downarrow}$) 
at the temperature $T=0$. In (a) we show a cut in  the $z=0$ plane
while (b) shows the column integrated signal. 
Parameters we used were  $q_x=0.25(\pi/d), q_y=q_z=0$, polarization
 $P=(n_{\uparrow}-n_{\downarrow})/(n_{\uparrow}+n_{\downarrow})=0.168$, $\Delta/(2J)=0.16$, and $U/(2J)=-1.86$. 
In figure (a) correlations vanish in the dotted areas.
We choose ${\bf r}+{\bf r' }-(\hbar t/m){\bf q}=0$ and
color-coding is again such that light colors imply high peaks and dark colors low.}
\label{fig:fig4}
\end{figure}

In  Fig.~\ref{fig:fig4} we show the FFLO-state noise correlations ($G_{\uparrow\downarrow}$) at zero temperature.
Fig.~\ref{fig:fig4} (a) shows the FFLO-state noise correlation in the $z=0$ plane,
and Fig.~\ref{fig:fig4} (b) demonstrates the integrated FFLO noise correlation.
As one can see, the Fermi surface shift creates asymmetry which remains also in the integrated signal.
Gapless regions of the FFLO-state are again reflected as areas without correlation peaks 
and this is clearly presented in Fig.~\ref{fig:fig4} (a). 

\section{Noise correlations at non-zero temperature}
\label{sec:non_0_T}

Until now we have assumed zero temperature, but now 
we generalize our computations to  non-zero temperatures. At non-zero temperatures it is useful to write 
the lowest energy 
state by using  fermionic quasiparticle operators 
$\hat \gamma_{\sigma,{\bf k},{\bf q}}$ and $\hat \gamma_{\sigma,{\bf k},{\bf q}}^{\dagger}$.
Creation and annihilation operators can be written in terms of these operators as
\beq
\label{eq:FFLO_bogo}
\begin{split}
&\hat c_{\uparrow,{\bf k}}=u_{{\bf k},{\bf q}}\hat \gamma_{\uparrow,{\bf k},{\bf q}}-v_{{\bf k},{\bf q}}\hat \gamma_{\downarrow,{\bf k},{\bf q}}^{\dagger}\\
&\hat c_{\downarrow,-{\bf k}+{\bf q}}=u_{{\bf k},{\bf q}}\hat \gamma_{\downarrow,{\bf k},{\bf q}}+v_{{\bf k},{\bf q}}\hat \gamma_{\uparrow,{\bf k},{\bf q}}^{\dagger} .
\end{split}
\enq
Now the lowest energy state ansatz is given by 
\beq
\label{eq:quasiGS}
|\Psi_{GS}\rangle =\prod_{\bf k}\hat \gamma_{\uparrow,{\bf k},{\bf q}}^{\dagger}\hat \gamma_{\downarrow,{\bf k},{\bf q}}^{\dagger}|0\rangle
\enq
and the lowest energy state 
is an ideal gas of quasiparticles. In other words the state in 
Eq.~\eqref{eq:quasiGS} is the vacuum state for these quasiparticle operators.

The non-vanishing 
two-operator expectation values in the densities in Eq.~\eqref{eq:fourierdens} at  non-zero temperatures  are
\beq
\label{eq:BCS_BP_dens_nonzero}
\begin{split}
&\langle \hat c_{\uparrow,{\bf k}}^{\dagger} \hat c_{\uparrow,{\bf k'}}\rangle =
\delta_{\bf kk'}(|u_{{\bf k},{\bf q}}|^2f(E_{\uparrow,{\bf k},{\bf q}})+|v_{{\bf k},{\bf q}}|^2(1-f(E_{\downarrow,{\bf k},{\bf q}})))\\
&\langle \hat c_{\downarrow,-{\bf k}+{\bf q}}^{\dagger} \hat c_{\downarrow,-{\bf k'}+{\bf q}}\rangle =
\delta_{\bf kk'}(|u_{{\bf k},{\bf q}}|^2f(E_{\downarrow,{\bf k},{\bf q}})+|v_{{\bf k},{\bf q}}|^2(1-f(E_{\uparrow,{\bf k},{\bf q}}))),
\end{split}
\enq
where the Fermi-Dirac distribution is given by \[ f(E)=\frac{1}{e^{E/(k_BT)}+1}.\]

while non-vanishing four-operator expectation values in the noise correlations in Eq.~\eqref{eq:fouriercor} at  non-zero temperatures  are given by
\beq
\label{eq:four_exp_val_nonzero}
\begin{split}
&\langle \hat c_{\uparrow,{\bf k}}^{\dagger}\hat c_{\downarrow,-{\bf k'}+{\bf q}}^{\dagger}\hat c_{\downarrow,-{\bf k'}+{\bf q }}\hat c_{\uparrow,{\bf k}}\rangle=
|u_{{\bf k},{\bf q}}|^2|u_{{\bf k'},{\bf q}}|^2f(E_{\uparrow,{\bf k},{\bf q}})f(E_{\downarrow,{\bf k'},{\bf q}})\\
&+|u_{{\bf k},{\bf q}}|^2|v_{{\bf k'},{\bf q}}|^2f(E_{\uparrow,{\bf k},{\bf q}})(1-f(E_{\uparrow,{\bf k'},{\bf q}}))+
|u_{{\bf k'},{\bf q}}|^2|v_{{\bf k},{\bf q}}|^2f(E_{\downarrow,{\bf k'},{\bf q}})(1-f(E_{\downarrow,{\bf k},{\bf q}})) \\
&+|v_{{\bf k'},{\bf q}}|^2|v_{{\bf k},{\bf q}}|^2(1-f(E_{\uparrow,{\bf k'},{\bf q}}))(1-f(E_{\downarrow,{\bf k},{\bf q}}))\\
&+\delta_{\bf kk'}\{|u_{{\bf k},{\bf q}}|^2|v_{{\bf k},{\bf q}}|^2[f(E_{\uparrow,{\bf k},{\bf q}})f(E_{\downarrow,{\bf k},{\bf q}})+
(1-f(E_{\uparrow,{\bf k},{\bf q}}))(1-f(E_{\downarrow,{\bf k},{\bf q}}))]\}\\
&\langle \hat c_{\uparrow,{\bf k}}^{\dagger}\hat c_{\uparrow,{\bf k'}}^{\dagger}\hat c_{\uparrow,{\bf k' }}\hat c_{\uparrow,{\bf k}}\rangle=
|u_{{\bf k},{\bf q}}|^2|u_{{\bf k'},{\bf q}}|^2f(E_{\uparrow,{\bf k},{\bf q}})f(E_{\uparrow,{\bf k'},{\bf q}})\\
&+|u_{{\bf k},{\bf q}}|^2|v_{{\bf k'},{\bf q}}|^2f(E_{\uparrow,{\bf k},{\bf q}})(1-f(E_{\downarrow,{\bf k'},{\bf q}}))+|u_{{\bf k'},{\bf q}}|^2|v_{{\bf k},{\bf q}}|^2
f(E_{\uparrow,{\bf k'},{\bf q}})(1-f(E_{\downarrow,{\bf k},{\bf q}}))\\
&+|v_{{\bf k'},{\bf q}}|^2|v_{{\bf k},{\bf q}}|^2(1-f(E_{\downarrow,{\bf k'},{\bf q}}))(1-f(E_{\downarrow,{\bf k},{\bf q}}))\\
&+\delta_{\bf kk'}\{|u_{{\bf k},{\bf q}}|^2|v_{{\bf k},{\bf q}}|^2[f(E_{\uparrow,{\bf k},{\bf q}})f(E_{\downarrow,{\bf k},{\bf q}})+(1-f(E_{\uparrow,{\bf k},{\bf q}}))
(1-f(E_{\downarrow,{\bf k},{\bf q}}))]\\
&-|u_{{\bf k},{\bf q}}|^2f(E_{\uparrow,{\bf k},{\bf q}})-|v_{{\bf k},{\bf q}}|^2(1-f(E_{\downarrow,{\bf k},{\bf q}}))\}.
\end{split}
\enq
Because the densities again cancel out most of the terms, the noise correlations turn out to be quite simple
\beq
\label{eq:non_zero_corr}
\begin{split}
&G_{\uparrow\downarrow}({\bf r},{\bf r'})=A(t)^4\sum_{\bf k}\bigg[\delta\left({\bf r}-\frac{\hbar t \tilde{\bf k}}{m}\right)
\delta\left({\bf r'}+\frac{\hbar t(\tilde{\bf k'}-{\bf q})}{m}\right)\\
&[f(E_{\uparrow,{\bf k},{\bf q}})f(E_{\downarrow,{\bf k},{\bf q}})+(1-f(E_{\uparrow,{\bf k},{\bf q}}))(1-f(E_{\downarrow,{\bf k},{\bf q}}))]|u_{{\bf k},{\bf q}}|^2|v_{{\bf k},{\bf q}}|^2\bigg]\\
&G_{\downarrow\uparrow}({\bf r},{\bf r'})=A(t)^4\sum_{\bf k}\bigg[\delta\left({\bf r}+\frac{\hbar t(\tilde{\bf k}-{\bf q}) }{m}\right)
\delta\left({\bf r'}-\frac{\hbar t \tilde{\bf k'}}{m}\right)\\
&[f(E_{\uparrow,{\bf k},{\bf q}})f(E_{\downarrow,{\bf k},{\bf q}})+(1-f(E_{\uparrow,{\bf k},{\bf q}}))(1-f(E_{\downarrow,{\bf k},{\bf q}}))]|u_{{\bf k},{\bf q}}|^2|v_{{\bf k},{\bf q}}|^2\bigg]\\
&G_{\uparrow\uparrow }({\bf r},{\bf r'})=A(t)^4\sum_{\bf k}\bigg[\delta\left({\bf r}-\frac{\hbar t \tilde{\bf k} }{m}\right)
\delta\left({\bf r'}-\frac{\hbar t \tilde{\bf k'} }{m}\right)\\
&[f(E_{\uparrow,{\bf k},{\bf q}})f(E_{\downarrow,{\bf k},{\bf q}})+(1-f(E_{\uparrow,{\bf k},{\bf q}}))(1-f(E_{\downarrow,{\bf k},{\bf q}}))]|u_{{\bf k},{\bf q}}|^2|v_{{\bf k},{\bf q}}|^2\\
&-|u_{{\bf k},{\bf q}}|^2f(E_{\uparrow ,{\bf k},{\bf q}})-|v_{{\bf k},{\bf q}}|^2(1-f(E_{\downarrow,{\bf k},{\bf q}}))\bigg]+
\delta({\bf r}-{\bf r'})\langle \hat n_{\uparrow}({\bf r},t)\rangle\\
&G_{\downarrow\downarrow }({\bf r},{\bf r'})=A(t)^4\sum_{\bf k}\bigg[\delta\left({\bf r}-\frac{\hbar t (\tilde{\bf k}-{\bf q}) }{m}\right)
\delta\left({\bf r'}-\frac{\hbar t (\tilde{\bf k'}-{\bf q}) }{m}\right)\\
&[f(E_{\uparrow,{\bf k},{\bf q}})f(E_{\downarrow,{\bf k},{\bf q}})+(1-f(E_{\uparrow,{\bf k},{\bf q}}))(1-f(E_{\downarrow,{\bf k},{\bf q}}))]|u_{{\bf k},{\bf q}}|^2|v_{{\bf k},{\bf q}}|^2\\
&-|u_{{\bf k},{\bf q}}|^2f(E_{\downarrow ,{\bf k},{\bf q}})-|v_{{\bf k},{\bf q}}|^2(1-f(E_{\uparrow,{\bf k},{\bf q}}))\bigg]+
\delta({\bf r}-{\bf r'})\langle \hat n_{\downarrow}({\bf r},t)\rangle.
\end{split}
\enq

At non-zero temperatures the state is the BCS-state if ${\bf q}=0$ and the densities are equal,
the BP-state if ${\bf q}=0$ and the densities are different. 
Since at finite temperatures the system can support polarization due to thermal effects
even if both quasiparticle dispersions are always positive,
we call the state gapless BP-state if ${\bf q}=0$ and one of the quasiparticle dispersions changes its sign.

\begin{figure}
\begin{tabular}{ll}
\includegraphics{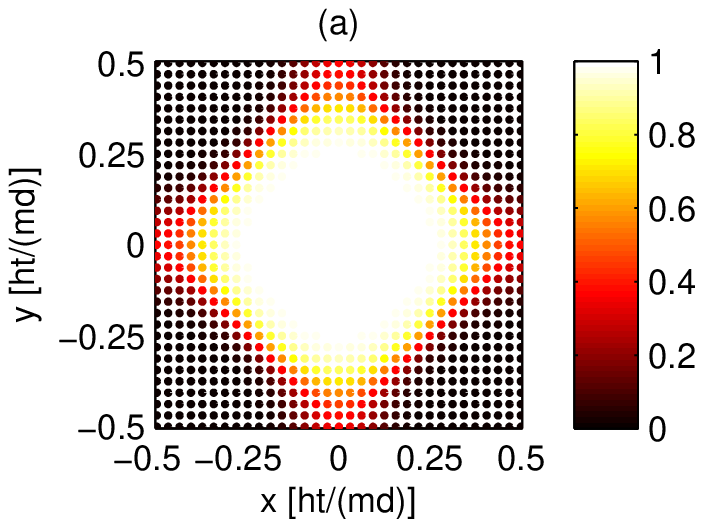} & \includegraphics{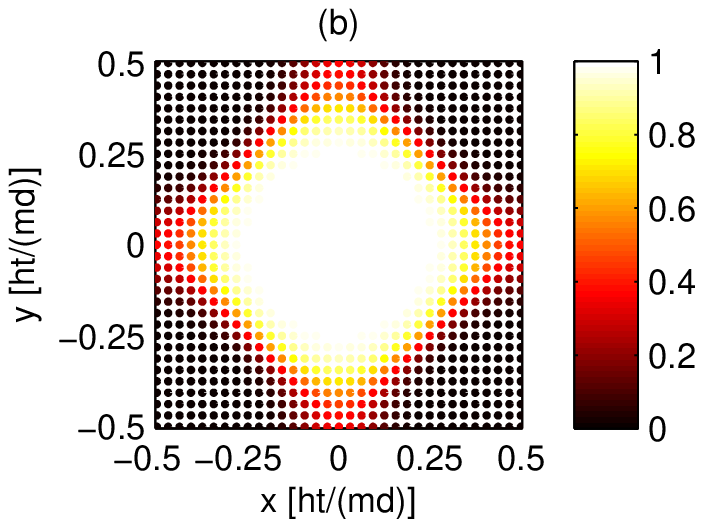} \\
\includegraphics{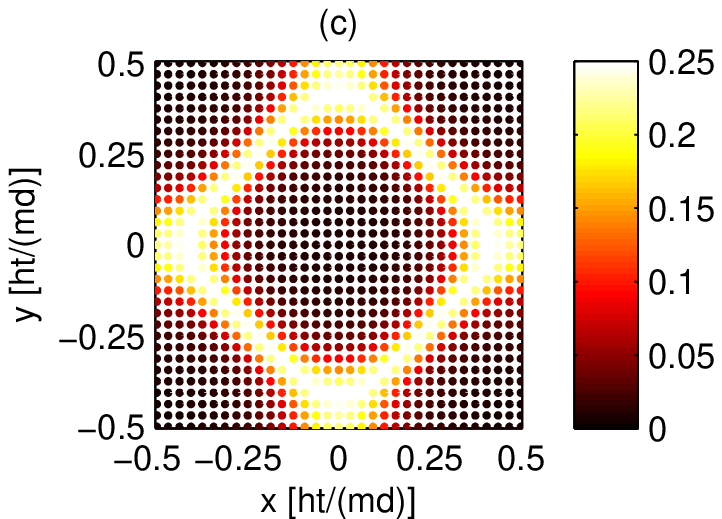} & \includegraphics{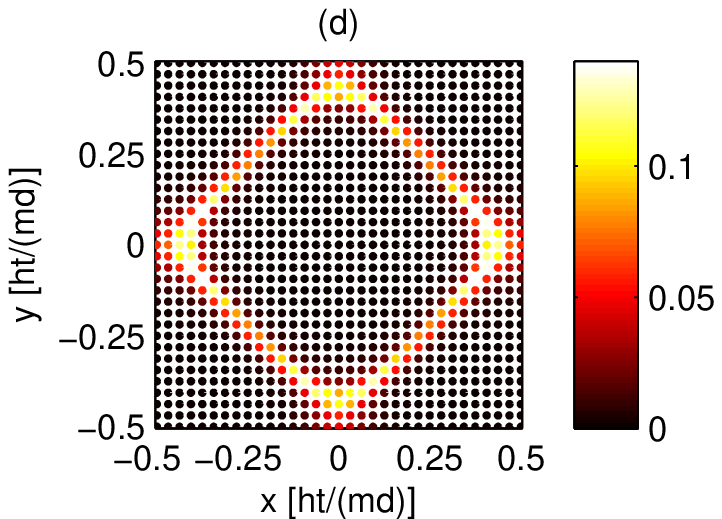}
\end{tabular}
\caption{[Color online] Figure (a) shows a cut of the BCS-state density in the $z=0$ plane when the temperature
is $k_BT/(2J)=0.196$ (just below $T_c$) while figure
(b) shows a cut of the normal-state density in the $z=0$ plane when the temperature
is $k_BT/(2J)=0.20$ (just above $T_c$). 
Figures (c) and (d) show the integrated BCS-state noise correlations between components 
at temperatures $k_BT/(2J)=0.0$ and $k_BT/(2J)=0.196$, 
respectively. 
The other parameters used were $n_{\uparrow}=n_{\downarrow}=0.20$, $U/(2J)=-1.86$, 
in figures (a) and (d) $\Delta/(2J)=0.09$ and in figure (c) 
$\Delta/(2J)=0.35$. Color-coding is again such that light colors imply high peaks and dark colors low.}
\label{fig:fig5}
\end{figure}

Fig.~\ref{fig:fig5}  demonstrates why the noise correlations are a better indicator 
of the superfluidity than densities and how the BCS-state
noise correlation between the components changes quite substantially when the temperature rises and the gap is reduced.
Fig.~\ref{fig:fig5} (a)  shows a cut of the BCS-state density in the $z=0$ plane when the temperature
is $k_BT/(2J)=0.196$ (i.e. just below the critical temperature). 
Fig.~\ref{fig:fig5} (b) shows a similar result just above the critical temperature. 
In Fig.~\ref{fig:fig5} (c) and (d) we compare the column integrated BCS-state noise correlations between the components at 
the temperatures $k_BT/(2J)=0.0$, $k_BT/(2J)=0.196$, respectively.
By comparing Fig.~\ref{fig:fig5} (a) and (b), one can see that the densities are almost 
the same for the BCS-state and the normal-state.
On the other hand in the normal-state the noise correlation is identically zero. 
Thermal effects are strongly present in the results in Fig.~\ref{fig:fig5} (d). This can be seen from the fact that
the correlations peak
in two areas, close to the center of the figure as well as near the Fermi surface. In the zero temperature
result in Fig.~\ref{fig:fig5} (c) 
there are high peaks only near the Fermi surface. The reason for this difference 
between low and higher temperatures is that when the temperature rises the peak heights around
the Fermi surface are reduced with reduced gap
while closer to the center the thermal effects are relatively weaker. 

\begin{figure}
\begin{tabular}{ll}
\includegraphics{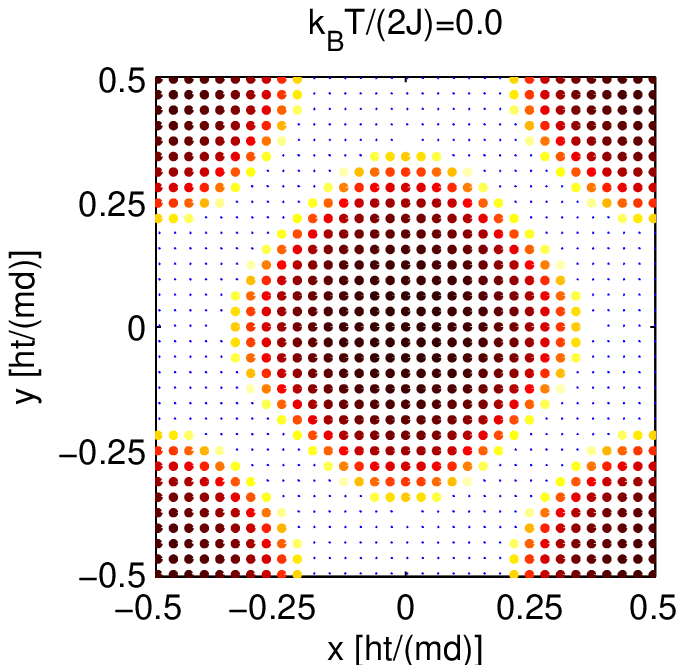} & \includegraphics{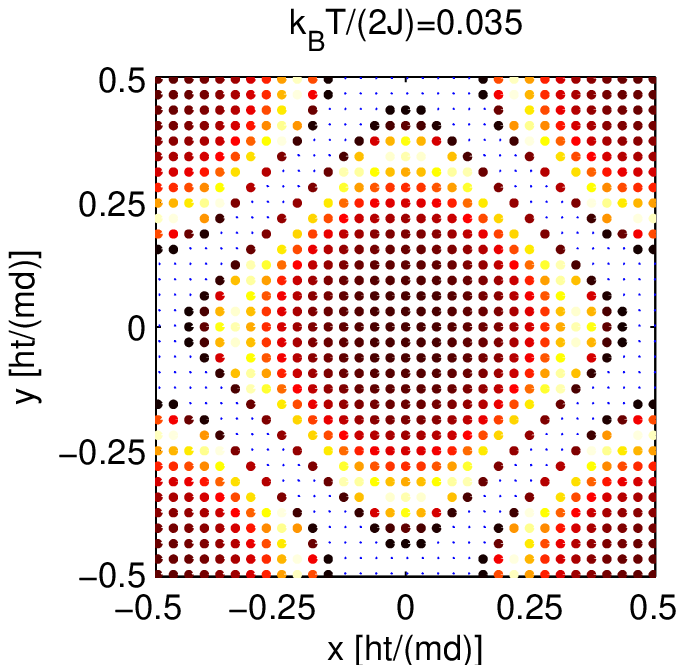} \\
\includegraphics{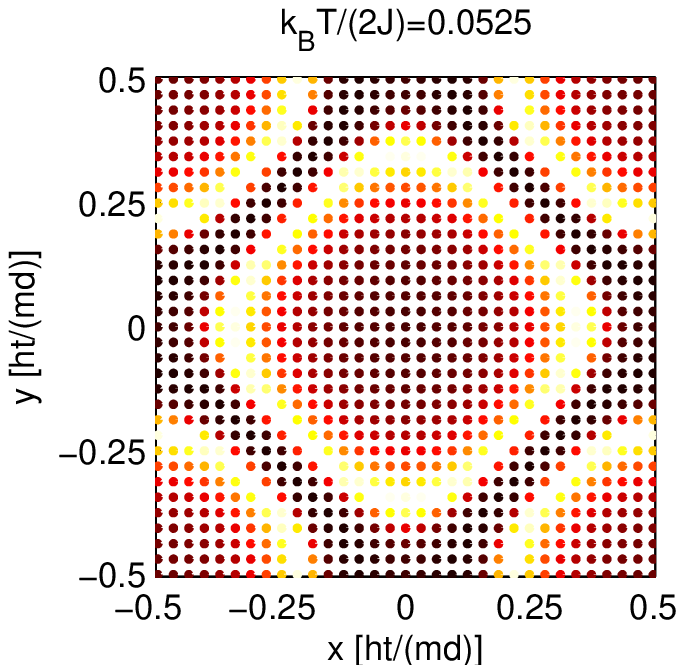} & \includegraphics{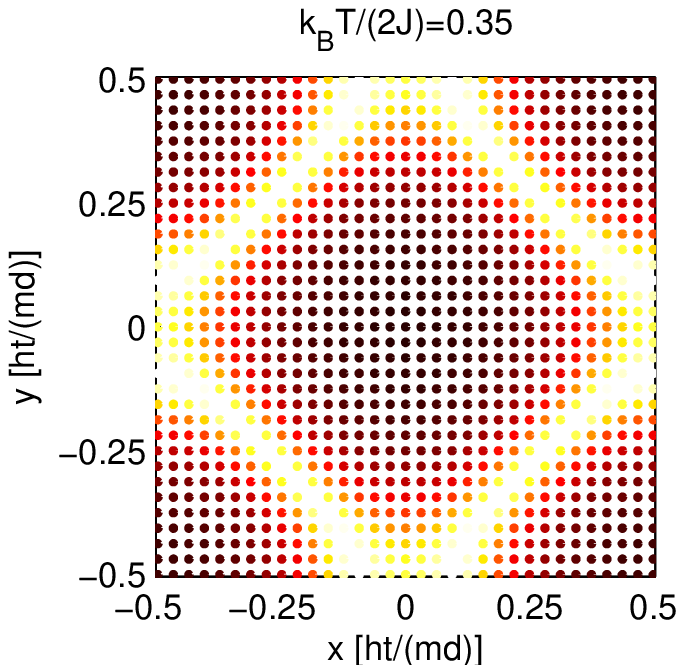}
\end{tabular}
\caption{[Color online] This figure 
demonstrates how temperature contributes to the noise correlations of the gapless BP-state.
In these figures the temperature gradually increases, while other things remain  equal. The last
figure corresponds to the minimum of the grand potential with polarization 
$P=(n_{\uparrow}-n_{\downarrow})/(n_{\uparrow}+n_{\downarrow})=0.40$, average 
filling fraction $(n_\uparrow+n_\downarrow )/2=0.30$, $\Delta/(2J)=0.318$ and coupling strength $U/(2J)=-3.0$.
We have chosen ${\bf r}+{\bf r' }=0$.
All figures show the cut in the $z=0$ plane.
At the upper row the dotted areas are peakless.
Color-coding is such that light colors imply high peaks and
dark colors low.}
\label{fig:fig6}
\end{figure}

Fig.~\ref{fig:fig6} demonstrates how the temperature  contributes to the gapless BP-state superfluid noise correlations.
As one can see, when temperature rises the peakless areas become smaller and eventually vanish. 
The reason for this is in the temperature fluctuations, which enables the system to carry polarization even with gapped 
dispersions.
These figures show also that the peakless areas vanish only gradually, when the temperature becomes non-zero.

\begin{figure}
\begin{tabular}{ll}
\includegraphics{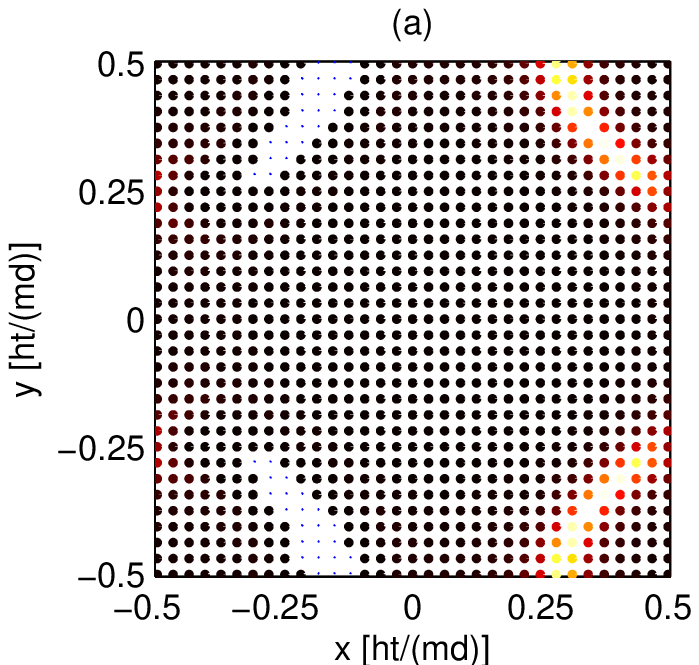} & \includegraphics{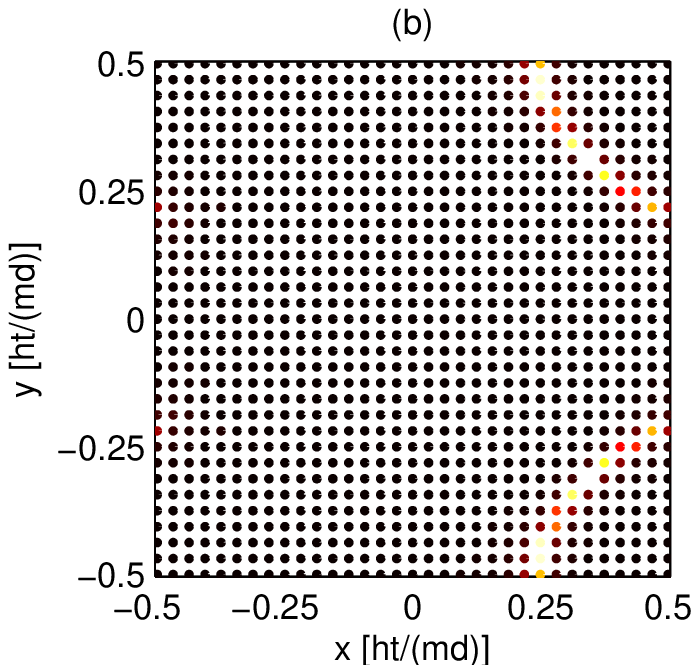} \\
\includegraphics{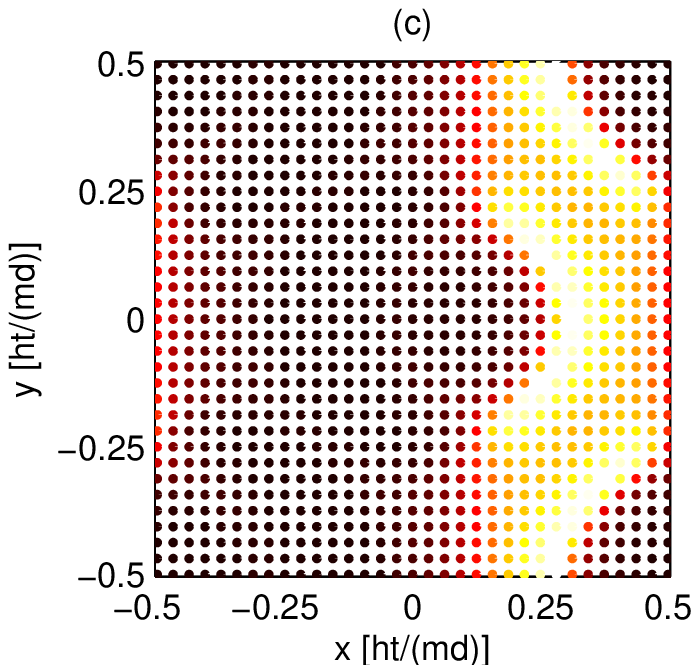} & \includegraphics{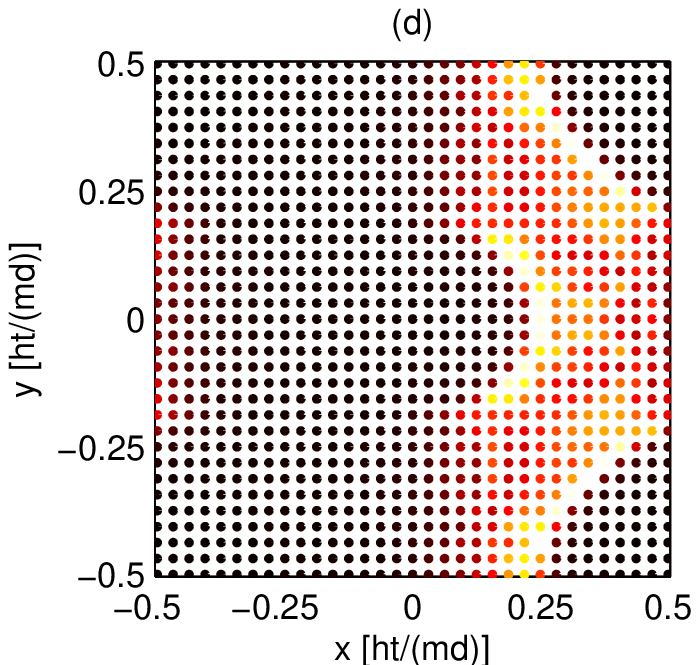}
\end{tabular}
\caption{[Color online] Finite temperature noise correlations ($G_{\uparrow\downarrow}$) of the FFLO state.
Figures (a) and (b) show cuts in the $z=0$ plane, while figures (c) and (d) are the column integrated  correlations. 
Figures (a) and (c) were calculated at $k_BT/(2J)=0.0475$ and figures (b) and (d)
were calculated $k_BT/(2J)=0.095$. In all figures we used the coupling strength $U/(2J)=-1.86$. 
In figures (a) and (c) $P=0.208$, $(n_\uparrow+n_\downarrow )/2=0.46$, $q_x=0.26(\pi/d)$, $q_y=q_z=0$, and $\Delta/(2J)=0.14$.
In figures (b) and (d) $P=0.24$, $(n_\uparrow+n_\downarrow )/2=0.37$, $q_x=0.21(\pi/d)$, $q_y=q_z=0$, and $\Delta/(2J)=0.046$.
We have chosen ${\bf r}+{\bf r' }-(\hbar t/m){\bf q}=0$.
In figure (a) correlations vanish in the dotted areas.
Color-coding is again such that light colors imply high peaks and
dark colors low.}
\label{fig:fig7}
\end{figure}

Finally, Fig.~\ref{fig:fig7}  demonstrates how temperature influences 
the noise correlations of the FFLO state. The states correspond to minima of the
free energy at their respective temperatures.
Figs.~\ref{fig:fig7} (a) and (b) show an example of what happens to 
the correlation between components on the $z=0$ plane 
when temperature rises from $k_BT/(2J)=0.0475$ to $k_BT/(2J)=0.095$. Figs.~\ref{fig:fig7} (c) and (d) 
demonstrate the same for the column integrated noise correlations. As one can see from the figures,
sharp areas without correlation peaks again disappear with increasing temperatures. 
However, the shift in the positions of the peaks 
persists even at non-zero temperatures.

It might not be easy to detect the gapless states via noise correlations because at the temperature regime 
where the  gapless BP-state becomes the lowest energy state, the noise correlations of the
gapless BP-state and non-gapless BP-state appear quite similar. 
On the other hand, by monitoring the peak heights at $z=0$ (and ${\bf r'}=-{\bf r}$), one can identify the gapless states.
The reason for this is that in the gapless BP-state the maximum peak height is always smaller than $A(t)^4/8$ 
while in the-non gapless state the maximum is always 
bigger than this. This can be seen from the fact that the term
$|u_{\bf k}|^2|v_{\bf k}|^2$ peaks on the Fermi surface corresponding to the average
chemical potential. On the other hand, near this surface the peak heights are
\[\begin{split}
&|u_{\bf k}|^2|v_{\bf k}|^2[f(E_{\uparrow,{\bf k}})f(E_{\downarrow,{\bf k}})+(1-f(E_{\uparrow,{\bf k}}))(1-f(E_{\downarrow,{\bf k}}))]\\
&\approx \frac{1}{4}[f(\Delta+\delta\mu/2)f(\Delta-\delta\mu/2)+(1-f(\Delta-\delta\mu/2))(1-f(\Delta+\delta\mu/2))],
\end{split}
\]
where $\delta\mu=\mu_{\uparrow}-\mu_{\downarrow}$ and then the term 
in the square brackets
is bigger than $1/2$, when $\Delta\pm \delta\mu/2>0$ (and the state is gapped)
and otherwise always lower than $1/2$.

\begin{figure}
\begin{tabular}{ll}
\includegraphics[width=0.490\columnwidth]{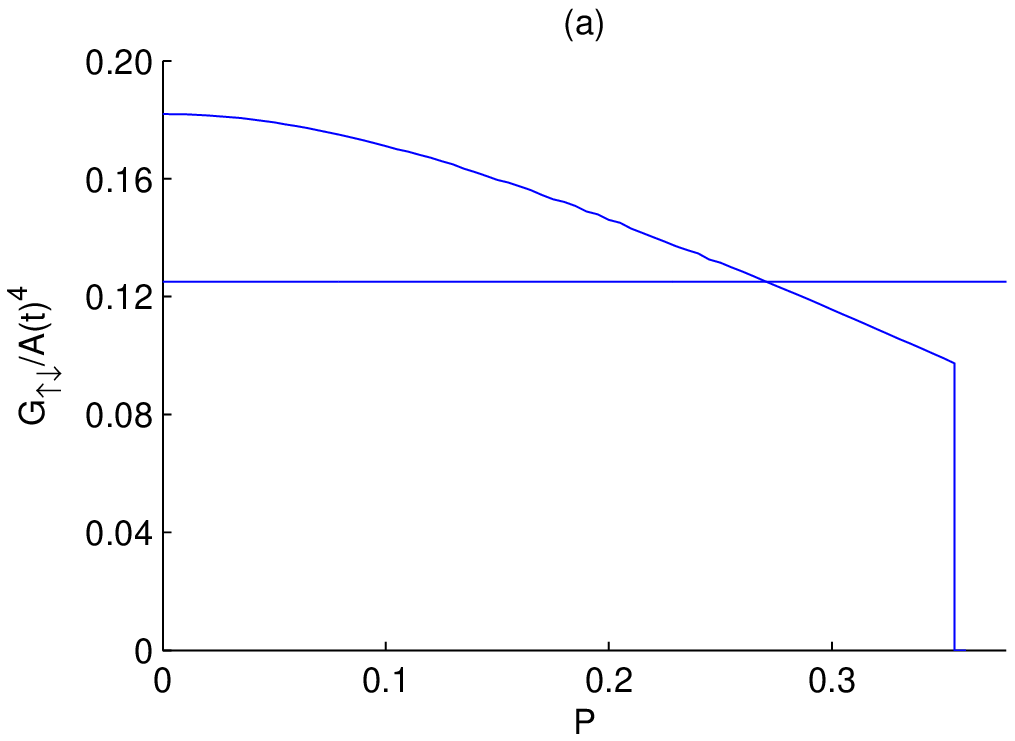} & \includegraphics[width=0.490\columnwidth]{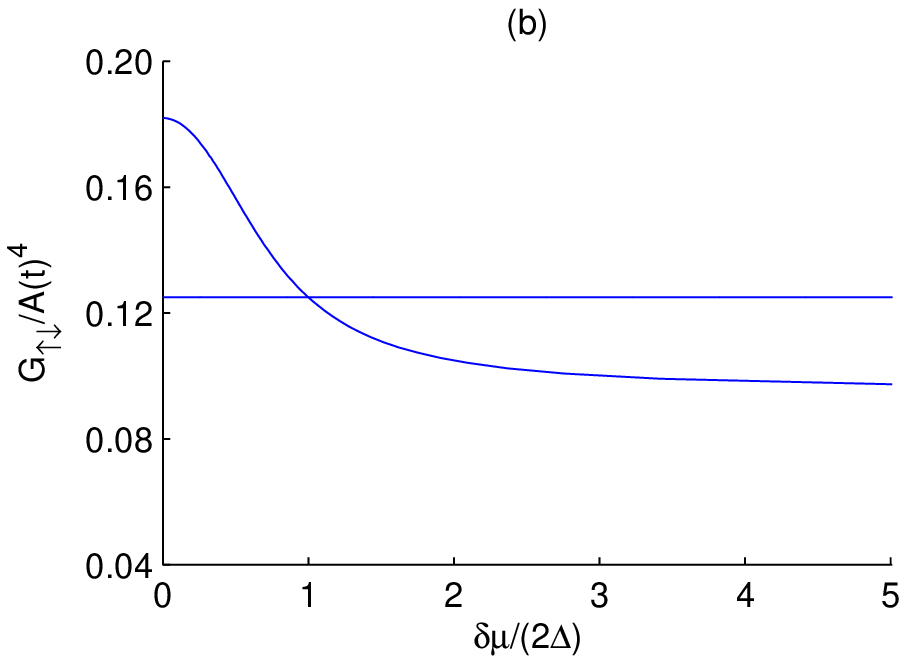} 
\end{tabular}
\caption{[Color online] This figure demonstrates how the polarization contributes to the maximum peak height.
(a) shows the maximum peak height as a function of the polarization
while (b) shows the maximum peak height as a function of the chemical potential difference
$(\mu_{\uparrow}-\mu_{\downarrow})/(2\Delta)=\delta\mu/(2\Delta)$. We used the parameters $k_BT/(2J)=0.45$ and an average 
filling fraction $(n_{\uparrow}+n_{\downarrow})/2=0.30$. 
We have canceled the time dependence of the maximum peak height  by dividing the maximum peak height by the 
scaling factor $A(t)^4$. The horizontal lines show value $1/8$.}
\label{fig:fig8}
\end{figure}

In Fig.~\ref{fig:fig8} we show the height of the maximum 
of peak height as a function of the polarization (Fig.~\ref{fig:fig8} (a)) and as a
function of chemical potential difference  (Fig.~\ref{fig:fig8} (b)).
Fig.~\ref{fig:fig8} (a) shows that when the polarization $P$ rises then the maximum peak height becomes smaller
and drops suddenly to zero when the gas becomes normal. With our 
parameters this happens when the  polarization is about $0.35$. By inspecting
Fig.~\ref{fig:fig8} (b) it is clear that when the BP-state is gapless,
i. e. when $\delta\mu>2\Delta$, the maximum peak height is indeed smaller than $1/8$, a result
which was made plausible above.

\section{Noise correlations in a one-dimensional lattice}
\label{sec:1D}


We call the lattice one-dimensional (1D) when $J_x\gg J_y=J_z$. In other
words a 1D lattice can be realized with  
three dimensional (3D) lattice with a very strong confinement in two
directions and weaker confinement in the third one.
Since the FFLO type states are more favorable in 1D than they are in 3D~
\cite{Machida1984a,Liu2007a,Parish2007a,Koponen2007b}
and the differences between different phases
can be more pronounced in 1D than in 3D, we now consider the noise
correlations in 1D lattices.
We assume that the gas in a 1D lattice is released only one direction
(in our case $x$-direction).
The released gas creates a set of almost identical one dimensional
tubes.

\begin{figure}
\includegraphics{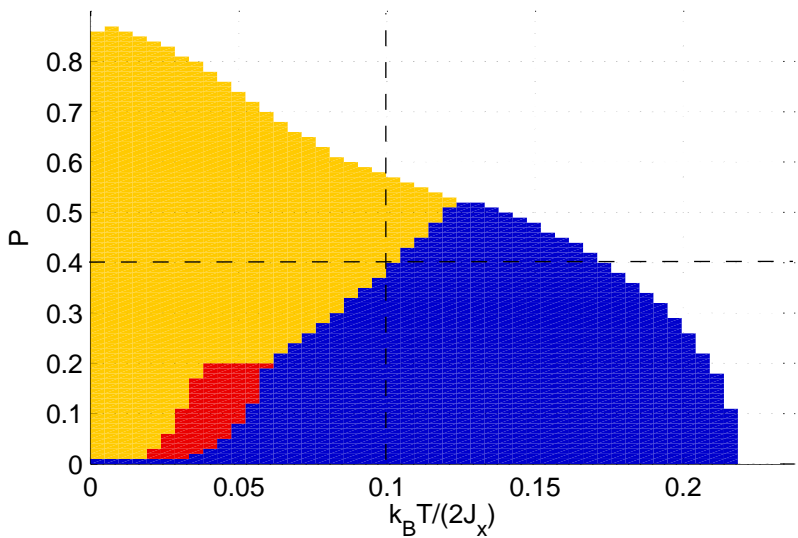}
\caption{[Color online] The phase diagram of Fermi gas in an effectively 1D-lattice~\cite{Koponen2007b}.
Colors (or shading): BCS/PB states=blue or dark gray, FFLO=yellow or light gray, phase separation=red or gray, and normal gas=white.
The parameters where such that 
the average filling fraction $(n_{\uparrow}+n_{\downarrow})/2=0.2$, $J_x=0.07E_r$, and $U=-0.2E_r$.
The dashed lines show the values $P=0.40$ and $k_BT/(2J_x)=0.104$ used in Fig.~\ref{fig:fig10} and in Fig.~\ref{fig:fig11}}
\label{fig:1D}
\end{figure}

Fig.~\ref{fig:1D} shows an example of 
the phase diagram in a 1D lattice~\cite{Koponen2007b}. It is clear that the FFLO region is remarkably
large and occupies the entire superfluid region at $T=0$. 
On the other hand the expected region for phase separation between the normal gas and the BCS-state (red) is 
dramatically smaller than in 3D~\cite{Koponen2007a}.
When the temperature rises there are phase transitions between the FFLO-state and, depending on polarization,
phase separation, the BCS/BP states, or the normal gas phases.
On Fig~\ref{fig:1D} we also denote by dashed lines those values of temperature and 
polarization we have used in our following noise correlation computations.

\begin{figure}
\begin{tabular}{ll}
\includegraphics{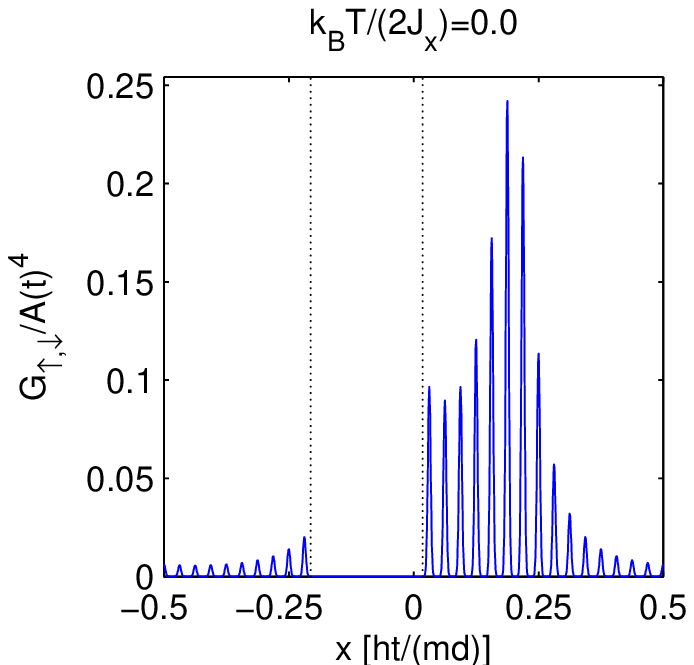} & \includegraphics{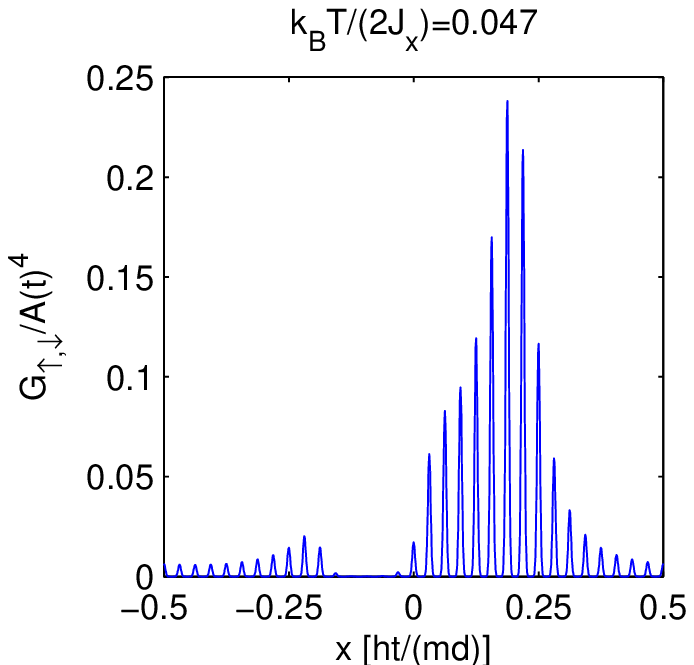} \\
\includegraphics{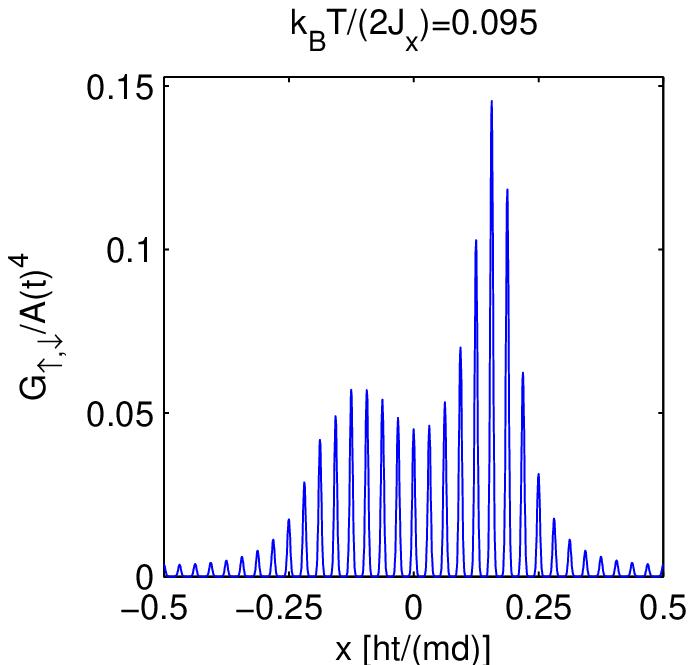} & \includegraphics{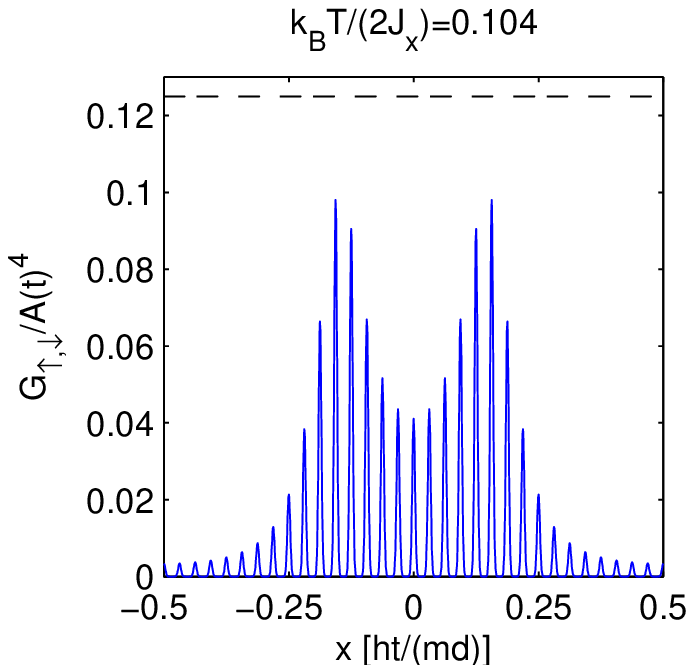}
\end{tabular}
\caption{[Color online] This figure demonstrates how temperature contributes to the noise correlations in a 1D lattice.
In these figures we gradually increase the temperature while keeping a polarization $P=0.40$ and an average filling fraction
$(n_{\uparrow}+n_{\downarrow})/2=0.20$ fixed. Figures in the upper row and bottom left show the FFLO-state noise correlations.
The figure in bottom right shows the gapless BP-state noise correlation.
In the first figure the dashed lines show the gapless region and  in the bottom right figure the dashed line shows the value
$1/8$. We choose $x+x'+\hbar t q_x/m=0$ in all these figures. In the BP-state $q_x=0$.
}
\label{fig:fig10}
\end{figure}

Fig.~\ref{fig:fig10} demonstrates how temperature contributes to the noise correlations in a 1D lattice. 
As is clear,
when the temperature becomes non-zero the gapless region does not vanish suddenly. Also, there is a phase transition
between the FFLO-state and the gapless BP-state when the temperature is (with our parameters)
between  $k_BT/(2J_x)=0.095$ and $k_BT/(2J_x)=0.104$. One can identify
this phase transition from the fact that in the FFLO-state the noise correlation is not symmetric with respect to $x=0$, while the 
gapless BP-state noise correlation is symmetric.
For the two-mode FFLO-state we expect a mirror image of the FFLO-state noise correlation 
shown here to appear. This mirror image would correspond to correlations with the pair momenta $-q_x$.
The figure which has been calculated at highest temperature ($k_BT/(2J_x)=0.104$)
shows that the gapless BP-state noise correlation is again always below $1/8$.

\begin{figure}
\begin{tabular}{ll}
\includegraphics{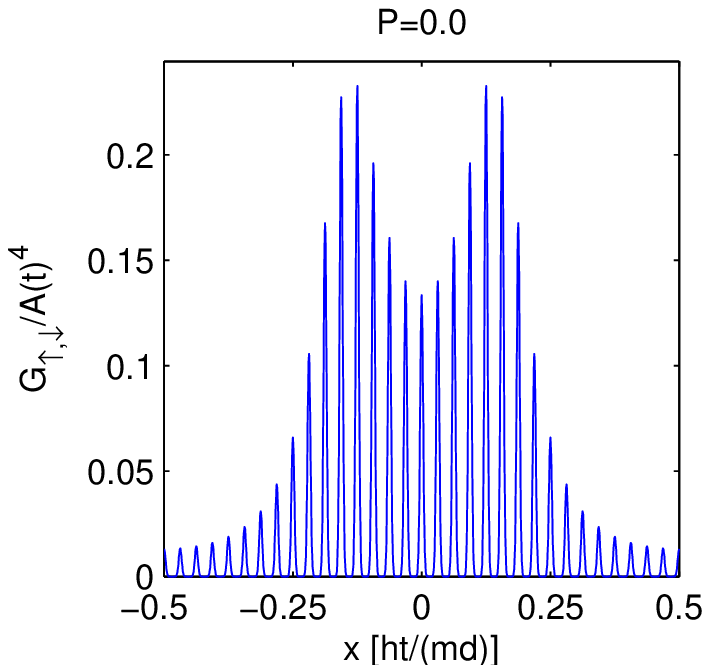} & \includegraphics{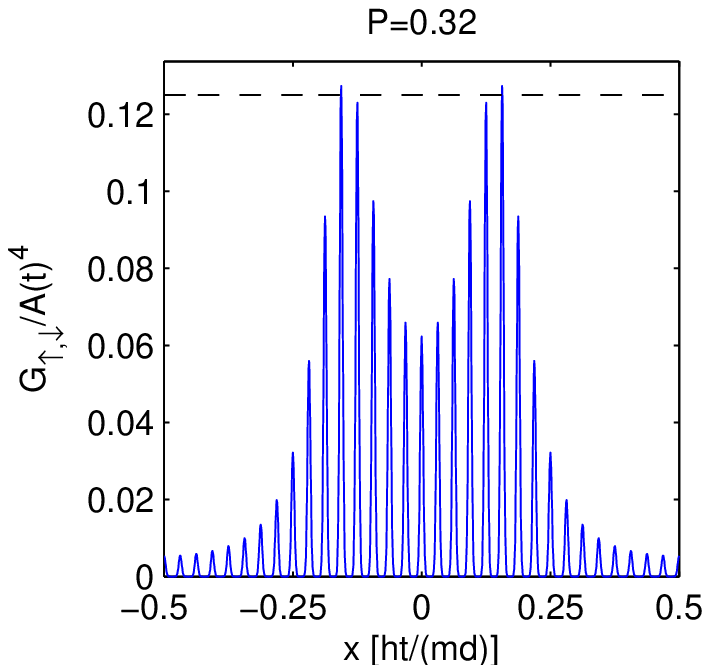} \\
\includegraphics{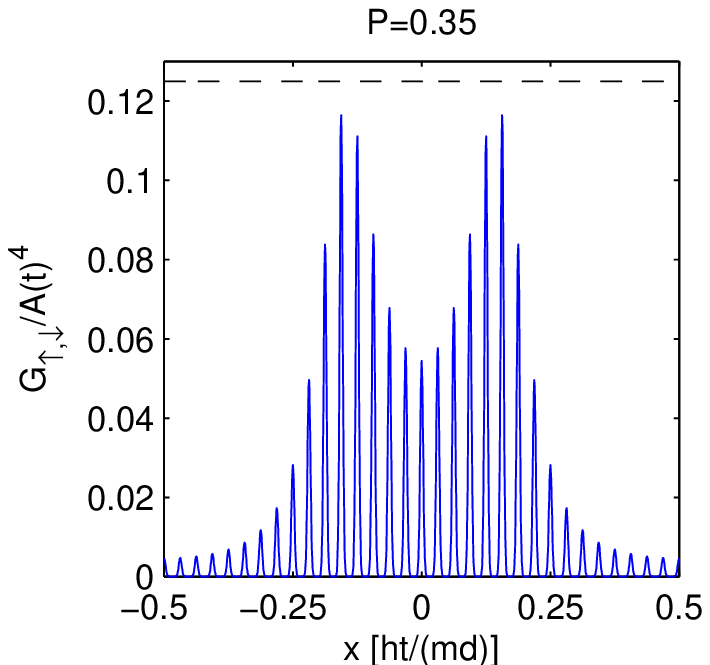} & \includegraphics{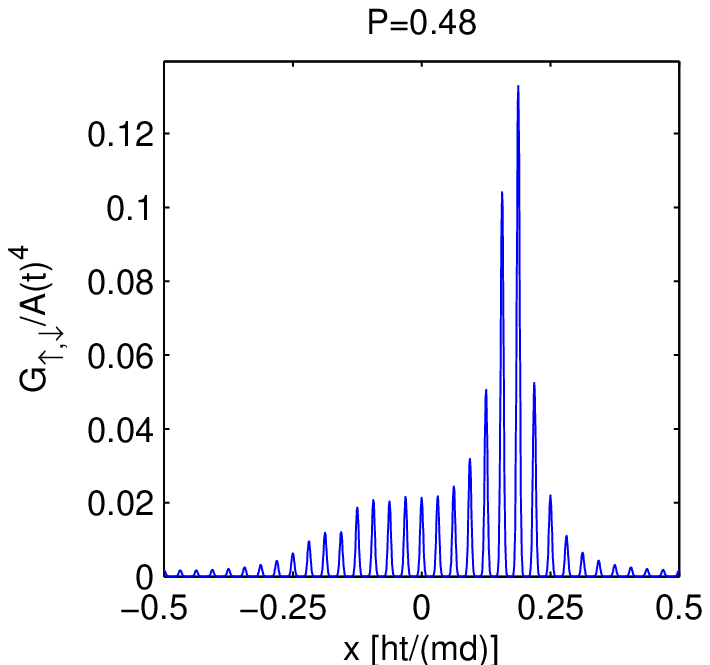}
\end{tabular}
\caption{[Color online] This figure shows how polarization contributes to the noise correlations in a 1D lattice.
Here we gradually increase the polarization while keeping the temperature fixed at $k_BT/(2J_x)=0.104$ and 
an average filling fraction at
$(n_{\uparrow}+n_{\downarrow})/2=0.20$. The upper left figure shows 
the BCS-state noise correlation and upper right figure shows the gapped BP-state
noise correlations. The bottom left figure shows the gapless BP-state noise correlation and
the figure on the bottom right is the FFLO-state noise correlation.
The dashed line shows the value $1/8$. 
We choose $x+x'+\hbar t q_x/m=0$ in all these figures. In the BCS- and the BP-states $q_x=0$.}
\label{fig:fig11}
\end{figure}

In Fig.~\ref{fig:fig11} we show how the polarization contributes to the noise correlations in a 1D lattice.
When the polarization is between $P=0.32$ and $P=0.35$ the non-gapless state becomes
the gapless and as one can see that the maximum of the gapped BP-state noise correlation is always bigger than
$1/8$. When the polarization is between $P=0.35$ to $P=0.48$, there is
a phase transition from the BP-superfluid state to
the FFLO-state which is visible in the pronounced asymmetry.

In one-dimensional systems, exact solutions for the problem of  
interacting Fermions exist~\cite{Gaudin1967a,Yang1967a,Lieb1968a}. 
Moreover, density matrix  
renormalization group (DMRG) calculations can be applied in one  
dimension to provide exact numerical results. It is known that mean  
field results deviate considerably from the exact ones in one  
dimension. In Ref.~\cite{Marsiglio1997a}, mean field  
calculations within the one-dimensional Hubbard model (which is the  
case we consider) were compared to the exact solutions~\cite{Gaudin1967a,Yang1967a,Lieb1968a}.
The ground state energies were found to be very similar, deviating at  
maximum a few percent. However, the mean field calculations were found  
to overestimate the excitation gap energy considerably, typically by a  
factor of 1.5-3. The deviations are at their strongest at half  
filling. These results imply that while the low temperature structure  
in the phase diagram of figure 9 is not likely to be affected by the  
mean field approximation, the critical temperatures are probably  
higher than given by the exact solution. The dominance of the FFLO  
state at low temperatures and for a large range of polarizations, as  
shown by the mean field results in Figure 9, is firmly established  
also by recent exact~\cite{Orso2007a,Hu2007a} and  
DMRG numerical studies~\cite{Feiguin2007a,Luscher2007a}. 
Note that the reference~\cite{Luscher2007a} considers also noise  
correlations. It is interesting that our results for
the FFLO correlations at zero temperature are quite similar with the  
more accurate computations
using DMRG by L\"{u}scher et al.~\cite{Luscher2007a}. In particular, both results show  
an area in
momentum space where correlation peaks are absent as well as  
pronounced asymmetry. In
the single mode FFLO ansatz the signals from both $\pm q$ are naturally  
missing while DMRG
computation is done for a trapped geometry and does show strong  
correlations corresponding to both $\pm q$. In summary, our results on the  
structure of the noise correlations for different states in 1D are  
essentially not affected by the use of mean field approximation, only  
the precise values of temperatures where these states occur are to be  
determined more carefully.

\section{Conclusions}
\label{sec:conc}

In this paper we have presented, at the mean field level, the noise correlations of 
the two component Fermi gas in an optical lattice.
We have shown that the noise correlations are a promising way to detect different 
phases in optical lattices. 
The different superfluid phases (BCS, FFLO, and BP) can 
be distinguished via the noise correlations and by mapping the correlations more
extensively, a ''Fourier analysis'' of the multimode FFLO-state is possible and could
be used to reveal the structure of the more complicated periodic order parameters.
We computed the noise correlations also at non-zero temperatures and  demonstrated 
that the differences between the correlations of different states
can persist at finite temperatures and some qualitative features are insensitive to temperature.
Also, regions of gapless quasi-particle dispersions can be visible in the 
noise correlation signals.

Other probes also exist. By letting the gas expand freely, one can
measure the momentum distribution by simply imaging the density
distribution
of the expanded cloud. From this momentum distribution one can, in
principle,
infer some important properties of the system. For example, one could
detect
the gapless BP-state in this way~\cite{Yi2006c}. However, this method
does not
appear to be a promising way to detect modulated phases~\cite{Luscher2007a}.
Spectroscopic means can also be considered. For example, using
two-photon Bragg spectroscopy~\cite{Stenger1999a,Vogels2002a} one can
probe the systems response
at specific momenta and in that way gain information on the
quasi-particle
excitations~\cite{vanOosten2005a,Challis2006a,Bruun2006a}. We are not
aware
of anyone analyzing this problem for fermions in optical lattices.
RF-spectroscopy has been proposed~\cite{Kinnunen2006a} for detecting 
a spatially nonuniform FFLO order parameter: an additional spectral
peak appears due to quasiparticles at the nodes of the order parameter. 
One very appealing property of noise correlation signals is their
sensitivity to
interactions and pairing effects. If pairing does not take place, noise
correlations
between different components vanish. This is in contrast to other
methods
where the influence of the other component is less direct.
Noise correlations can be especially useful in distinguishing between  
possible pseudo-gap effects and superfluidity. Pseudo-gap implies  
non-condensed pairs, then noise correlations between momenta ${\bf k}+{\bf q}$ and  
$-{\bf k}+{\bf q}$ appear not only for a single or a few discrete values of ${\bf q}$, but  
for a thermally distributed set of ${\bf q}$ values. The appearance of  
correlations only for a certain ${\bf q}$ signals condensation of the pairs.

Density-density correlations of a one-dimensional system were also discussed
quite extensively and were shown to contain very clear information on the
structure of the pair-wavefunction as well as on the quasi-particle dispersions.
In one-dimensional systems, order parameter modulations are favored and
possible problems with signals becoming smoothed out
by column integration can be avoided. For this reason they are very attractive
candidates to observe FFLO-type states experimentally. However, how
to balance the requirements of sufficiently reduced dimensionality 
with sufficient phase coherence~\cite{Parish2007a}
is to a large extent still unclear.

It is physically possible that in a three-dimensional system 
a phase-separation occurs between the normal state and 
the BCS-state if the system is polarized~\cite{Bedaque2003a,Martikainen2006a,Haque2006a}.
This means that the normal state  exists in  one part of the lattice and the BCS-state exists  
in another part of the lattice,
typically in the core if a harmonic trap potential is included. However, other more exotic 
possibilities also exist~\cite{Koponen2007b}.
Interestingly, the phase-separation between the normal gas and a paired-state could
be visible in the noise correlations between components. This follows from
the fact that the noise correlations between components in the normal state
vanish, whereas in the paired state the correlations are at their strongest around the
Fermi momentum. In a lattice superimposed by a trap the local density, and therefore
the local Fermi momentum, is different in different areas of the gas. This may allow to
identify spatial phase separation and shell structures of normal and paired states from
the freely expanded cloud where momentum has been mapped into position.
 

{\it Acknowledgments}
This work was supported by the National Graduate School in
Materials Physics, Academy of Finland (Projects
Nos. 115020, 213362, 121157, 207083) and conducted as a part of a
EURYI scheme award. See www.esf.org/euryi.

\bibliographystyle{apsrev}
\bibliography{bibli}

\end{document}